\documentclass[fleqn,usenatbib]{mnras}

\usepackage{float}

\usepackage{newtxtext,newtxmath}

\usepackage{xcolor}
\usepackage{anyfontsize}

\usepackage[T1]{fontenc}

\usepackage{pifont}

\usepackage{hyperref}
\hypersetup{colorlinks=true, citecolor=blue, linkcolor=blue}

\DeclareRobustCommand{\VAN}[3]{#2}
\let\VANthebibliography\thebibliography
\def\thebibliography{\DeclareRobustCommand{\VAN}[3]{##3}\VANthebibliography}

\usepackage{graphicx}	
\usepackage{amsmath}	
\title[Black hole in a near pristine galaxy]{A black hole in a near pristine galaxy 700 million years after the Big Bang}
\author[Maiolino et al.]{Roberto Maiolino,$^{1,2,3}$ \thanks{E-mail: \href{mailto:rm665@cam.ac.uk}{rm665@cam.ac.uk}}
Hannah \"{U}bler,$^{4}$
Francesco D'Eugenio,$^{1,2}$
Jan Scholtz,$^{1,2}$
\newauthor
Ignas Juod\v{z}balis,$^{1,2}$
Xihan Ji,$^{1,2}$
Michele Perna,$^{5}$
Volker Bromm,$^{6,7}$
Pratika Dayal,$^{8}$
  \newauthor
Sophie Koudmani,$^{9,10,11,12}$
Boyuan Liu,$^{10,13}$
Raffaella Schneider$^{14,15,16,17}$
Debora Sijacki,$^{1,10}$
\newauthor
Rosa Valiante,$^{15,16}$
Alessandro Trinca,$^{15,16,18}$
Saiyang Zhang,$^{6,7}$
Marta Volonteri,$^{19}$
\newauthor
Kohei Inayoshi,$^{20}$
Stefano Carniani,$^{21}$
Kimihiko Nakajima,$^{22}$
Yuki Isobe,$^{1,2,23}$
Joris Witstok,$^{24,25}$
\newauthor
Gareth C. Jones,$^{1,2}$
Sandro Tacchella,$^{1,2}$
Santiago Arribas,$^{5}$
Andrew Bunker,$^{26}$
\newauthor
Elisa Cataldi,$^{27,28,29}$,
Stephane Charlot,$^{19}$
Giovanni Cresci$^{29}$
Mirko Curti,$^{27}$
Andrew C. Fabian,$^{10}$
\newauthor
Harley Katz,$^{30,31}$
Nimisha Kumari,$^{32}$
Nicolas Laporte,$^{33}$
Giovanni Mazzolari,$^{4}$
\newauthor
Brant Robertson,$^{34}$
Fengwu Sun,$^{35}$
Bruno Rodriguez Del Pino,$^{5}$
Giacomo Venturi,$^{22}$
}
\begin{document}
\label{firstpage}
\pagerange{\pageref{firstpage}--\pageref{lastpage}}
\maketitle
%
\begin{abstract}
The recent discovery of a large number of massive black holes within the first two billion years after the Big Bang, as well as their peculiar properties,
        have been largely unexpected based on the extrapolation of the properties of luminous quasars.
        These findings have prompted the development of several theoretical models for the early formation and growth of black holes, which are, however, difficult to differentiate. We report the metallicity measurement around a gravitationally lensed massive black hole at redshift $7.04$ (classified as a Little Red Dot), hosted in a galaxy with very low dynamical mass. The weakness of the [OIII]5007 emission line relative to the narrow H$\beta$ emission indicates  extremely low metallicity, about $4\times 10^{-3}$ solar, and even more metal poor in the surrounding few 100 pc. We argue that such properties cannot be uncommon among accreting black holes around this early cosmic epoch. Explaining such a low chemical enrichment in a system that has developed a massive black hole is challenging for most theories. Models assuming heavy black hole seeds (such as Direct Collapse Black Holes) or super-Eddington accretion scenarios struggle to explain the observations, although they can potentially reproduce the observed properties in some cases. Models invoking ``primordial black holes'' (i.e. putative black holes formed shortly after the Big Bang) may potentially explain the low chemical enrichment associated with this black hole, although this class of models also requires further developments for proper testing.
\end{abstract}




\section{Introduction}
\label{sec:intro}

The James Webb Space Telescope (JWST) has opened a new window on the exploration of the formation and evolution of black holes in the early universe. Indeed, JWST's unprecedented sensitivity and wavelength coverage has enabled the detection of active galactic nuclei (AGN) at high redshift that have luminosities much lower than quasars discovered from ground based surveys, and also pushing the AGN discovery frontier to much higher redshifts \citep[e.g.][]{Kocevski_AGN,Ubler2023,
Harikane_AGN,Matthee2023,Maiolino_AGN, Maiolino2024, Kokorev2023_AGN,
Furtak2023_AGN,Greene2024,Taylor2024,Juodzbalis2024,Juodzbalis2025type1JADES,Scholtz2025_type2,Taylor2025_capersLRD,
Geris2025,Mazzolari2024CEERS,Tripodi2024_lowZ,Chisholm2024}. 
These AGN are typically identified through a broad component of the permitted lines (typically H$\alpha$ or H$\beta$), without a counterpart in [OIII] (hence ascribing them to the Broad Line Regions of AGN rather than outflows), as well as high excitation lines typical of the hard and intense radiation field of AGN. The low luminosity of these JWST-discovered AGN is partly due to lower black hole masses and/or lower accretion rates \citep{Maiolino_AGN,Juodzbalis2024,Juodzbalis2025type1JADES}. However, these lower luminosity AGN are not simply the high redshift version of AGN known locally or at intermediate redshifts, and neither are scaled down version of quasars at similarly high redshifts. The new population of AGN discovered by JWST seems to be quite different from the AGN populations previously known. Indeed, they are found to be extremely X-ray weak \citep{Maiolino_xray_weak,Ananna2024,Yue2024} and radio weak \citep{Mazzolari2024radio,Mazzolari2024CEERS}. They also typically show weak or no variability \citep{Kokubo2024,Zhang2025variability,Ji2025,Furtak2025variab,Naidu2025}. Most of them appear to be overmassive in terms of black hole to stellar mass ratio, relative to the local relation \citep{Harikane_AGN,Maiolino_AGN,Furtak2023_AGN,Juodzbalis2024,Juodzbalis2025type1JADES,Chen2025hostsLRD}. Although part of this overmassive nature is certainly due to selection effects \citep{Juodzbalis2025type1JADES,Li2024,Geris2025}, it is still remarkable that some of these AGN have black holes that are nearly as massive as their host galaxies. Many of these new AGN lack strong high ionization lines \citep[e.g.][]{Lambrides2024} as well as the classical iron emission bumps \citep{Trefoloni2024,Ji2025GNz11}. They are also exceptionally nitrogen-rich \citep{Isobe2025}.
Multiple models have been suggested to interpret these results, ranging from high (super-Eddington) accretion rates \citep{Madau2024,Pacucci2024superEdd,Schneider2023,King2025superEdd,Maiolino_xray_weak}, heavy obscuration by dust-poor or dust-free medium \citep{Maiolino_xray_weak,Ji2025,Ji2025localLRD}, heavy seeds \citep{Natarajan23},  seeding from merging star clusters \citep{Partmann2025,Rantala2025}, possibly resulting from ``feedback-free starbursts'' \citep{Dekel2025a,Dekel2025b} PopIII and PopIII.1 seeds \citep{Nandal2025_PopIII,Prole2025_PopIII,Sanati2025_PopIII1, Cammelli2025_PopIII1, Tan2005_PopIII1}, and
primordial black holes \citep{Zhang2025arXiv250317585Z,dayal2024_pbh,Dayal_Maiolino2025,Ziparo2025},
just to mention a few.

About 10\%-30\% \citep{Hainline2025} of the newly discovered broad line AGN have a peculiar V-shaped continuum, with blue UV slope and red optical slope. These have been dubbed ``Little Red Dots'' (LRDs) \citep{Matthee2024,Kocevski_AGN,Hviding2025LRDs}. They typically show absorption in the Balmer lines \citep{Matthee2024,Juodzbalis2024b,Lin2025LRDsgrism,DEugenio2025,Deugenio2025_restfrabs}, which indicates absorption by large columns of dense gas along the line of sight \citep{Juodzbalis2024b,Inayoshi2024,
Chang2025absorpt}, probably associated with absorption that might be responsible for the observed X-ray weakness. They are typically characterized by prominent (though often smooth) Balmer breaks; although these were initially interpreted as originating from evolved stellar populations, it has later been
shown that these are  connected with the Balmer absorption features and associated with dense gas absorption \citep{Inayoshi2024,Ji2025,DEugenio2025,Naidu2025,Degraaff25b_cliff}. LRDs are typically found at z$>$4 and steeply declining in number density at later cosmic epochs, although luminous examples have been found at z$\sim$2 \citep{Juodzbalis2024b,Wang2024,Loiacono2025,Euclid2025Bisigello} and, more recently, a few local counterparts have been identified \citep{Lin2025localLRD,Ji2025localLRD}.

One of the first discovered and most prototypical LRD is Abell2744-QSO1 (hereafter QSO1). This is a strongly lensed, triply imaged LRD, initially identified via NIRCam imaging in the field of the cluster Abell2744 \citep{Furtak2023} and then spectroscopically confirmed with MSA prism spectroscopy \citep{Furtak2023_AGN} to be at z=7.04.
 The lensed images are unresolved, with an upper limit on the (continuum) size of only 30~pc in radius. Low resolution spectroscopy also revealed broad H$\beta$, typical of type~1 AGN, a prominent Balmer break, and the V-shaped continuum typical of LRDs. Already based on the low resolution spectrum \cite{Furtak2023_AGN} indicated that this LRD hosts an AGN with a mass of $M_{BH}\sim 3\times 10^7~M_\odot$.
\cite{Ma2025QSO1} investigated further the spectral shape of the continuum of this object by highlighting the difficulty of reproducing it, especially its Balmer break, with simple stellar or AGN models. Their fiducial model ascribed the break to an evolved stellar population with a mass of $4\times 10^9~M_\odot$, which, given the extremely compact size, would imply an extremely high stellar surface density ($1.4\times 10^6~M_\odot~{\rm pc}^{-2}$), not seen in any other local or lower redshift galaxies with similar or higher masses. However, \cite{Ji2025} obtained NIRSpec-IFU data as part of the BlackTHUNDER programme, both with the prism and with the high resolution grating, revealing H$\beta$ absorption; they showed that both the Balmer break and H$\beta$ absorption can be fully explained in terms of extremely dense gas absorption along the line of sight. They also disentangled the narrow component of H$\beta$, as well as [OIII]5007, which are both spectrally unresolved; from these they obtain a tight upper limit on the dynamical mass of the host galaxy, which turns out to be an order of magnitude lower than the stellar mass inferred when assuming a stellar origin of the Balmer break -- this is further confirmation that the Balmer break is non-stellar in origin. Furthermore, both \cite{Ji2025} and \cite{Furtak2025variab} also detect variations of the EW(H$\beta$) of the three images of QSO1, indicative of source variability and, therefore, unambiguously confirming that QSO1 is hosting an accreting black hole and also the non-stellar nature of the optical continuum. \cite{DEugenio2025} extended the wavelength range of the BlackTHUNDER NIRSpec-IFU data to fully cover H$\alpha$, confirming the presence of prominent Balmer absorption and obtaining tighter constraints on the host galaxy's dynamical mass.

More recently, \cite{Juodzbalis2025_directBH} performed a more detailed analysis of the BlackTHUNDER data of QSO1, focusing on H$\alpha$, to trace the inner rotation curve of the system. They leverage the gravitational lensing sheer, together with independent analyses using spectroastrometry and detailed kinematic modelling \citep[using the MOKA3D environment][]{Marconcini2023}, to resolve the sphere of influence of the black hole. Their analysis has revealed a pure Keplerian rotation around a point mass of $5\times 10^7~M_\odot$; this direct determination of the BH, is fully consistent with the black hole mass estimation inferred from the broad H$\beta$ and using single-epoch virial relations calibrated locally \citep{Furtak2023_AGN,Ji2025}. Additionally, \cite{Juodzbalis2025_directBH} pointed out that the observed Keplerian rotation leaves little room for any stellar component, with an upper limit of $M_{star}<2\times 10^7~M_\odot$. Their constraint results into a tight lower limit on the black hole to stellar mass ratio of $M_{BH}/M_{star}>2$, which is the highest ever found.

Another interesting feature of the original prism spectrum of QSO1 is the faintness of the [OIII]5007 line. This, per se, would not necessarily be an anomaly among AGN, and may simply indicate a weak Narrow Line Region, which was found in the past for powerful quasars at high redshift \citep{Netzer2004noNLR}, although much more luminous than QSO1. However, the high resolution spectrum has shown a clear narrow component of both H$\beta$ and H$\alpha$, revealing that a Narrow Line Region, or ISM ionized by star formation, is actually present and that [OIII]5007 is actually anomalously weak relative to the narrow components of the hydrogen lines. In this paper we analyse the BlackTHUNDER high resolution data to assess the [OIII] weakness more thoroughly and more quantitatively, and show that it is tracing very low metallicity of the gas embedding the black hole. We also discuss the possible implications for our understanding for black hole seeding.

A flat $\Lambda$CDM cosmology is adopted throughout based on the latest results of the Planck collaboration \citep{Planck2020}, with $H_0 = 67.4 \, \mathrm{km \, s^{-1} \, Mpc^{-1}}$, $\Omega_\text{m} = 0.315$, $\Omega_\text{b} = 0.0492$.
At $z = 7.04$, 
an on-sky separation of $1\arcsec$ corresponds to $5.53 \, \text{physical kpc}$ (pkpc). The lensing magnification results into a stretching of this scale by a factor of about 3.5 \citep{Furtak2023_AGN}. 


\begin{figure*}
	\centering
\includegraphics[width=0.6\linewidth]{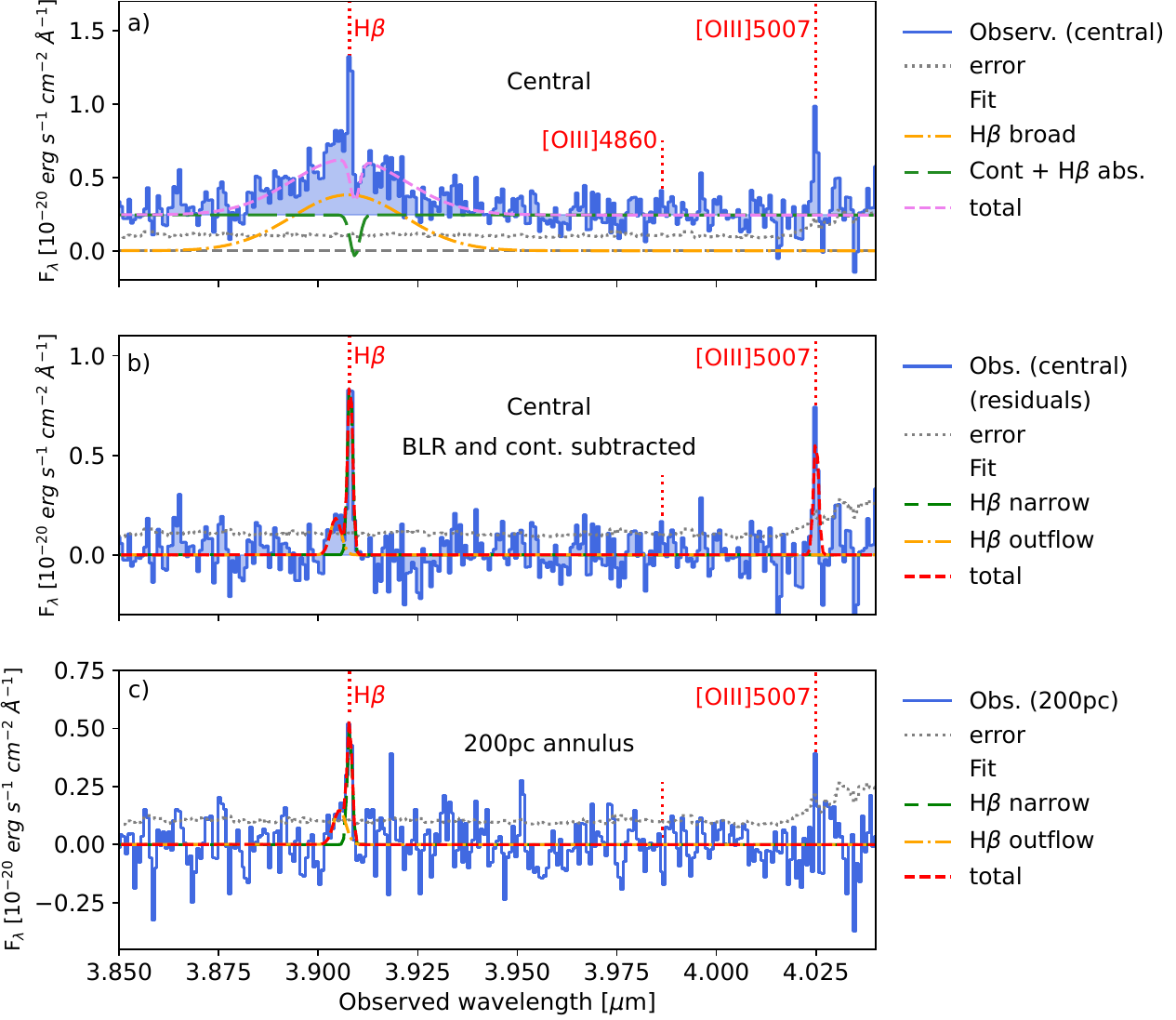}
	\caption{Spectra of QSO1 around H$\beta$ and [OIII]5007. a) Spectrum extracted from the central region (blue solid line), with the fitted broad component of H$\beta$ (orange dot-dashed line), continuum and H$\beta$ absorption (green long dashed line), as in \citet{Ji2025}, while the violet short-dashed line shows the total fit. b) The same central spectrum where the broad H$\beta$, continuum and H$\beta$ absorption have been subtracted to highlight the narrow component of H$\beta$ and [OIII]5007.  c) Spectrum extracted from a semi-annulus from 0.1$\arcsec$ ($\sim 150$~pc) to 0.2$''$ ($\sim 300$~pc) from the centre -- H$\beta$ is clearly detected while [OIII]5007 is formally undetected (flux mostly on a single pixel).
    In all panels, the dotted gray line shows the 1$\sigma$ error level. 
	}
\label{fig:spectra}
\end{figure*}

\section{Observations and data reduction}
\label{sec:datared}

We used the NIRSpec-IFS spectral cubes from the BlackTHUNDER programme PID-5015 (PIs H. \"{U}bler and R. Maiolino), already presented in \cite{Ji2025} and \cite{DEugenio2025}. The observations and data processing are extensively discussed in these two papers. Here we simply recall that the observations targeted image A of QSO1 (lensing magnification factor $\mu \sim 5.8-6.15$, depending on the adopted model, \citealt{Ji2025,Furtak2023_AGN}) with the NIRSpec IFU mode, both with the high resolution disperser G395H (for a total exposure of 7.4 hours) and with the low resolution prism (for a total of 2 hours), although the latter is not used in this paper.
We note that image B has slightly higher magnification but located near a bright foreground galaxy in projection, which makes its analysis more difficult, while image C has a significantly lower magnification, hence the choice of selecting image A for the IFU observation. As discussed more extensively in in \cite{Ji2025} and \cite{DEugenio2025}, the data were processed with the JWST pipeline, but with additional steps to correct for the $1/f$ noise and for outliers rejection. Before combination, the cubes were resampled to a scale of $0''.05$ per spaxel. As noted by previous studies, the ERROR extension of the NIRSpec IFS cubes underestimates the real error \citep{Ubler2023,Rodriguez2024GANIFS}; following the same procedure as previous works, the error extension has been rescaled to match the value measured in regions of the cube free of sources.


\begin{figure}
	\centering
\includegraphics[width=.6\linewidth]{"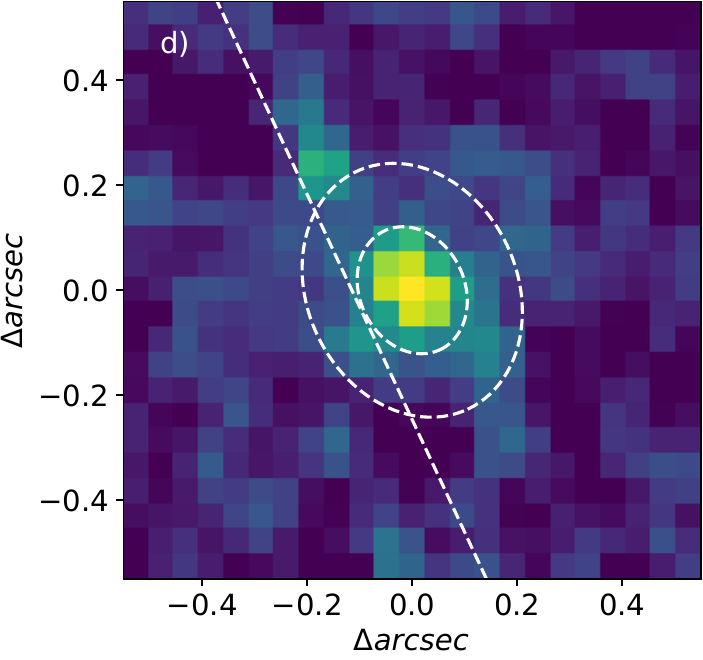"}
	\caption{Map of H$\beta$ narrow (obtained by simply collapsing the three central channel of the continuum-subtracted line) with overlaid the central aperture and annulus used for extracting the spectra in panels a,b and c of Fig.\ref{fig:spectra}, respectively; [OIII] falls in the detector gap in the region on the East (i.e. left) of the dashed straight line, hence this portion of the annulus was not used for extracting the spectrum.
	}
\label{fig:Hb_map}
\end{figure}

\section{Data analysis: Spectral fitting and spatial extension}
\label{sec:dataan}

Fig.\ref{fig:spectra}a shows the spectra extracted from the central 0.2$''$, i.e. from within the central r$<$150~pc.
The aperture is not perfectly circular in order to account for the shape of the PSF.
Unfortunately, the Point Spread Function (PSF) of the IFU observations is difficult to determine, as it does not  follow exactly the JWST theoretical PSF -- it also tends to be slightly elongated along the direction of the IFU's slices \citep{DEugenio2024}.
In the case of QSO1 we leverage the fact that the continuum is unresolved, based on the NIRCam images \citep{Furtak2023_AGN} and therefore we can take the continuum as a reference to trace the PSF. We find that the continuum between H$\beta$ and [OIII] (the spectral region of main interest), is well described by an ellipse with axial ratio 0.8 and major axis $PA=-25^\circ$. The central extraction aperture is shown by the central ellipse in Fig.\ref{fig:Hb_map}, overlaid on the map of the H$\beta$ narrow component (hereafter $H\beta _N$). Note that this map was conservatively created by simply collapsing the three spectral
channels (continuum-subtracted) around the peak H$\beta_N$. This implies that the map includes
also some contribution from the broad line, but this does not affect our
result as the map is not used for quantitative purposes.

As already shown in \cite{Ji2025} and \cite{DEugenio2025}, the central spectrum confirms the broad H$\beta$ associated with the AGN's BLR, with H$\beta$ absorpion \citep[visible at higher significance in H$\alpha$][]{DEugenio2025}. However, most importantly, it clearly reveals a narrow component of H$\beta$ and  extremely weak [OIII]5007 emission. As already mentioned in the introduction, such weak [OIII] emission was also noticed by \cite{Furtak2023_AGN} in the low resolution, prism spectrum, but they could not assess the intensity relative to the corresponding narrow component of H$\beta$, because the latter is blended with the broad component coming from the BLR in the prism spectrum.
In the new high resolution spectrum it is possible to spectrally resolve and disentangle the broad component of H$\beta$ ($FWHM\sim 2600~\rm{km/s}$) from the narrow component (which is spectrally unresolved, with $FWHM<75 \, \rm{km/s}$), as illustrated in Fig.\ref{fig:spectra}a. We adopt the same fitting approach as in \cite{Ji2025}, with Gaussian components (with lower limit widths given by the spectral resolution) to disentangle the narrow and broad H$\beta$ components, and which also includes an absorption component, and [OIII]5007. 
We could in principle also include  H$\alpha$ in the simultaneous fitting \citep[as in ][]{DEugenio2025}, but our focus is on the [OIII]/H$\beta$ ratio, while the different PSF at the wavelength of H$\alpha$, convolved with the extension of the emission (discussed later), could introduce artifacts -- however, reassuringly we obtain flux ratios that are fully consistent with those obtained by \cite{DEugenio2025}.
The spectrum extracted from the central aperture with radius of 0.1$''$, subtracted of the broad component and continuum, is shown in Fig.\ref{fig:spectra}b,  further highlighting the prominence of the narrow component of H$\beta$ (H$\beta_N$) and the weakness of [OIII]5007. H$\beta$ has a residual weak blue wing after subtracting the broad component, which (as discussed further below) is spatially resolved and probably associated with an outflow, which we fit with a separate Gaussian and whose flux is not included in H$\beta_N$.

\begin{table}
    \centering
    \begin{tabular}{lcc}
    \hline
    $F([OIII]5007)/F(H\beta_N)$ & (central) & $0.55\pm 0.16$ \\
    $F([OIII]5007)/F(H\beta_N)$) & (R$\sim$200) & $< 0.41$ \\
    $F([OIII]5007)/F([OII]3727)$ & (central) & $>3.1$ \\
    $F([OIII]4363)/F([OIII]5007)$ & (central) & $<0.30$ \\
    $F([SII]6717,6731)/F(H\alpha_N)$ & (central) & $<0.08$ \\
    $F([NII]6584)/F(H\alpha_N)^*$ &  (central) & $<0.29$ \\
    $F(H\alpha_N)/F(H\beta_N)^*$ &  (integrated) & $ 2.6^{+0.6} _{-0.5}$ \\
        \hline
    \end{tabular}
    \caption{Flux ratios for the narrow emission lines. 
    Notes: $^*$ From \citep{DEugenio2025}
    }
\label{tab:fluxes}
\end{table}

The inferred ratio between [OIII]5007 and H$\beta_N$ is extremely low, less than unity ([OIII]5007/H$\beta_N \sim 0.6$), as reported in Table \ref{tab:fluxes}. It should be noted that the Gaussian fit of [OIII]5007 does not capture the full profile of the line; this is because the narrow components are unresolved and poorly sampled by the NIRSpec spectral pixels. Low spectral sampling is also the reason why the peak of [OIII] relative to H$\beta _N$ appears somewhat higher than the $\sim 0.6$ ratio given by the flux ratio -- when lines are unresolved and poorly sampled, the intensity of the line peaks also depends on how the line flux is distributed on the few spectral pixels. However, we have checked that the simple integration of the flux in the spectral pixels covering the lines give the same flux ratios as the Gaussian fits.

As we will discuss further in Sect.~\ref{sec:lowZ}, such extremely low [OIII]/H$\beta_N$ ratio is
very rarely found among high-z galaxies and generally interpreted in terms of extremely low metallicity.

\begin{figure}
	\centering
    \includegraphics[width=0.7\linewidth]{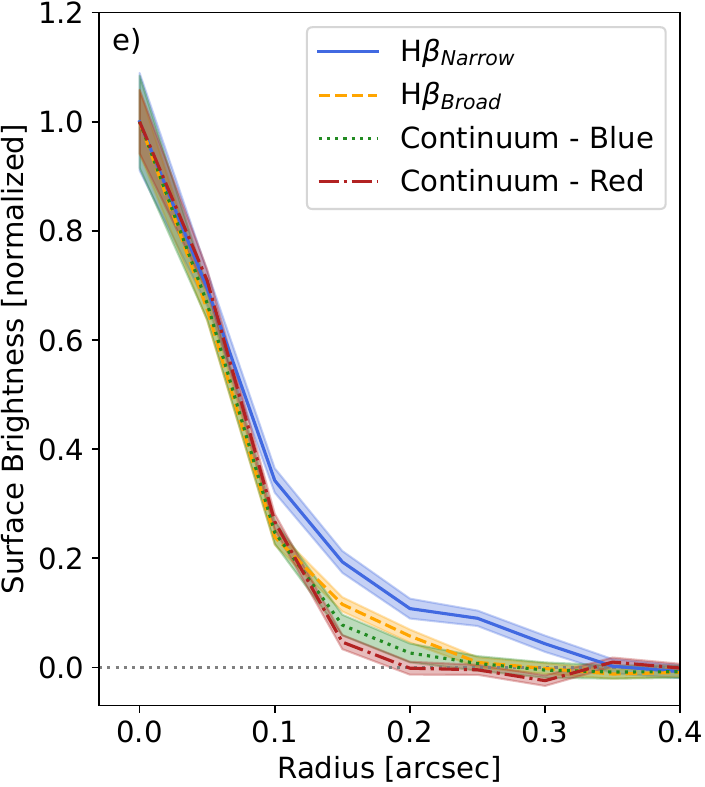}

	\caption{Radial profiles (normalized to the peak) of the narrow component of H$\beta$ (solid blue, the shaded area is 1$\sigma$ error on the mean), the broad component of H$\beta$ (dashed orange), the continuum on the blue side of H$\beta$ (dotted green) and the continuum on the red side of H$\beta$ (dot-dashed red), which trace the PSF. The narrow component has a compact core but it also shows a clear extended component. 
	}
\label{fig:profiles}
\end{figure}

As also pointed out by \cite{Juodzbalis2025_directBH}, who presented a spatially resolved analysis of H$\alpha$ in QSO1, the hydrogen lines have some spatial extension. While \cite{Juodzbalis2025_directBH} focused on H$\alpha$, here we focus on H$\beta$ as we need it in comparison with [OIII].
The H$\beta$ extension is noticeable in Fig.\ref{fig:profiles}, which shows the radial profile of H$\beta_N$ (blue solid line), compared with the radial profile of the broad component of H$\beta$ and the continuum.
For simplicity and to avoid degeneracies, the radial profile of H$\beta_N$ shown in Fig.\ref{fig:profiles} is obtained by simply collapsing the three spectral channels around the peak of $H\beta _N$; this is conservative as it implies that the actual radial profile of $H\beta _N$ is probably even more extended (as it includes also some contribution from the broad line). As already mentioned,
given that the continuum is not resolved in the NIRCam images, it can be used to trace the PSF. The profile of the continuum blueward of $H\beta$ was obtained by collapsing the spectral channels in the region 3.700--3.835~$\mu$m (green dashed line), and the profile of the continuum redward of H$\beta$ was obtained by collapsing the spectral channels in the region 3.944--3.700~$\mu$m (red dot-dashed line).
The profile of broad $H\beta$ (dashed orange line) is extracted from the high S/N region between 3.894--3.904~$\mu$m; this  also includes part of the ``blue wing'', which we tentatively ascribe to a low metallicity outflow, on the grounds that this profile is slightly more extended than the continuum and also seen in the outer annulus, which is discussed further below. The comparison of the profiles 
illustrates that, while a large fraction H$\beta_N$ comes from a very compact, unresolved region, it also has a component extending on scales of $\sim 0.2''$ ($\sim 300$~pc deprojected). The broad component of H$\beta$ has some indication of extension, likely because it includes the weak blue wing (with smaller width relative to the broad component) that is probably tracing an outflow.

Motivated by the observed extension of H$\beta_N$, in Fig.\ref{fig:spectra}c we also show the spectrum extracted from a semi-annulus with radius between $0.1''$ and $0.2''$ (150-300~pc) from the centre; note that we had to exclude the South-West part of the annulus because of [OIII] falling in the detectors gap in that region (this is the region to the SE of the dashed line in Fig.\ref{fig:Hb_map}).
The contamination from the spectrum in the central region (due to the wings of the central PSF) to the  spectrum extracted from the annulus is estimated to be below 15\%; this is confirmed by the fact that the
the broad H$\beta$ is undetected in the spectrum of the annulus. There is a weak blue wing, which may originate from a mild outflow, as discussed above. Yet, the most important finding is that  H$\beta_N$ is still clearly detected. There is only a weak hint of [OIII]5007, which is formally undetected (the putative flux is only on one single spectral pixel, much narrower than the line spread function). The inferred upper limit of the narrow line ratio in this annulus is even lower than in the central region: [OIII]/H$\beta_N<0.41$ ($3\sigma$). Note that this is a conservative upper limit - some contamination from the central spectrum (where the [OIII]/H$\beta_N$ ratio is higher) to the narrow lines would imply that the upper limit on the [OIII]/H$\beta_N$ ratio in the annulus is  even lower.

\begin{figure}
	\centering
    \includegraphics[width=\linewidth]{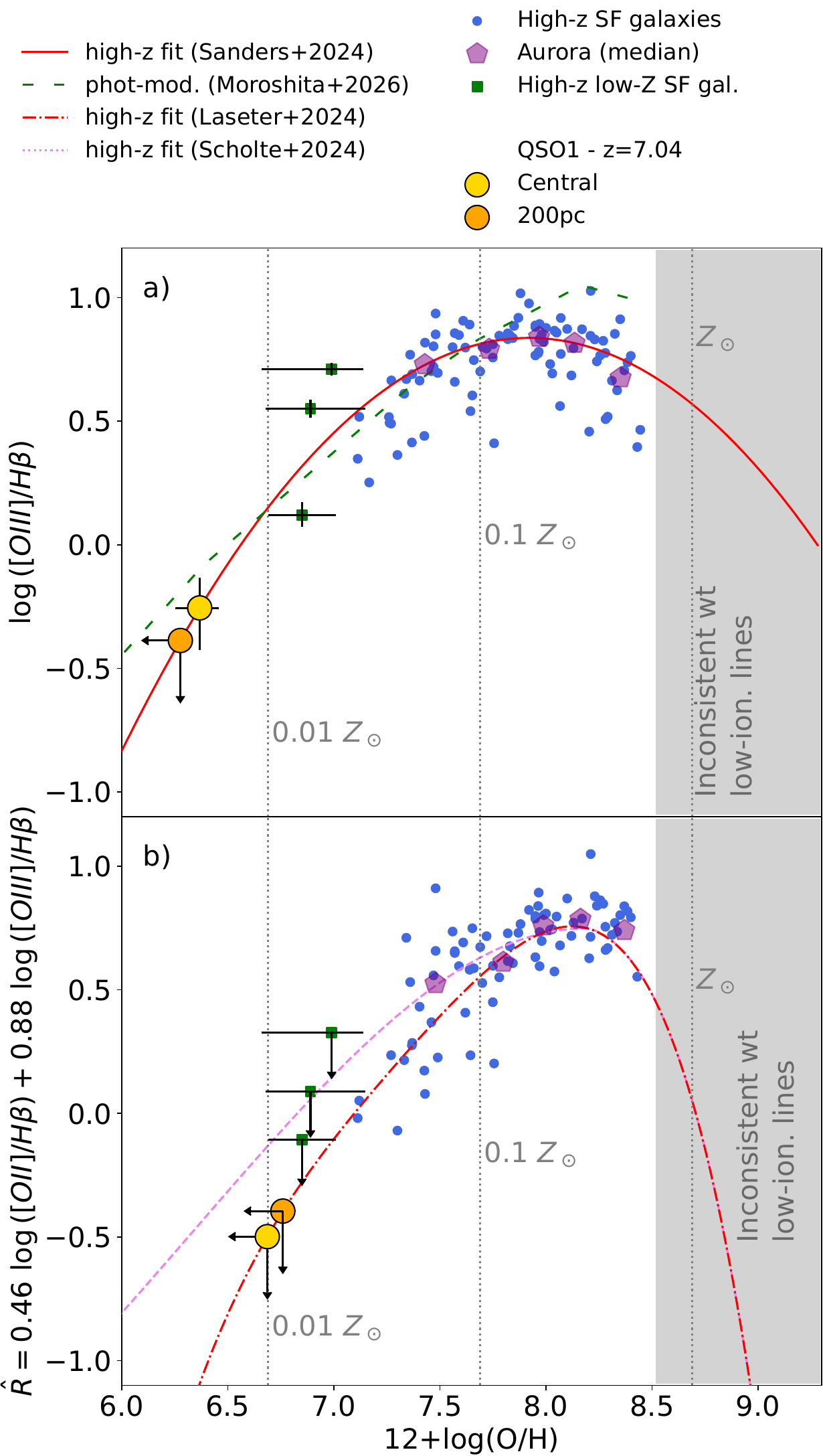}
	\caption{
    Metallicity constraints on QSO1 inferred from the observed emission line ratios. 
    (a) [OIII]5007/H$\beta$ vs 12+log(O/H); (b) $\hat{R}=0.46~R_2+0.88~R_3$ vs 12+log(O/H)
    The red solid line shows the  calibration for [OIII]/H$\beta$ from \citet{Sanders2024}
    and while the red dot-dashed line shows the calibration for $\hat{R}$ from \citet{Laseter2024metallicitycalib}, in both cases using high-z star forming galaxies. Small blue points are individual, T$_e$-based measurements at high-z from \citet{Cataldi2025Marta}. Green squares are other high-z galaxies at low metallicity  \citep{Willott2025,Cullen2025,Mowla2024}. Purple pentagons are median values at high redshift from the Aurora survey \citep{Sanders2025Aurora}. The QSO1 line ratios are shown with large circles, yellow and orange for the central region and for the annulus at $\sim$200~pc, respectively.
    The green long-dashed line in (a) is the photoionization models relation by \citet{Nakajima2022} adopted by \citet{Morishita2025_lowZ} and
    short-dashed purple line in (b) is the calibration for $\hat{R}$ from \citet{Scholte2025}; both these alternative calibrations would give even lower metallicities. 
    The shaded area indicate the high metallicity solutions excluded by the low ionization lines (Methods). In both panels errobars are at 1$\sigma$ while upper limits are at 3$\sigma$
	}
	\label{fig:met}
\end{figure}


\section{Evidence for very low metallicity}
\label{sec:lowZ}

As already mentioned, the extremely low [OIII]/H$\beta_N$ ratio is
very rarely found among high-$z$ galaxies \citep{Vanzella2023,Curti2023,Hsiao2025_lowZ,Morishita2025_lowZ,Nakajima2025_lowZ,Vanzella2025_lowZ} and, in these low mass systems,  is explained in terms of extremely low metallicity \citep{
Nakajima2022Empress,Nakajima2022}.
In Appendice \ref{app:high_density} and \ref{app:low_ion} we discuss that, unlike the H$\beta$ broad (from the BLR), alternative scenarios for explaining the low [OIII]/H$\beta$  (such as extremely high density of the ISM or low ionization) are physically implausible. 

As discussed in \cite{DEugenio2025}, given the extremely narrow width of [OIII], H$\beta_N$ and H$\alpha_N$, these components are likely powered by weak star formation in the host. We therefore infer the metallicity by using the calibrations recently derived for high-$z$ star forming galaxies. However, in Appendix \ref{app:NLR} we show that the results would hold, and would actually be reinforced, also assuming photoionization by the AGN. Fig.~\ref{fig:met}a shows the location of QSO1
on the [OIII]/H$\beta$ versus metallicity diagram,  calibrated by \cite{Sanders2024}  (red solid line). For comparison 
we show (with small blue points) direct, individual metallicity measurements of
high-$z$ galaxies collected by \cite{Cataldi2025Marta}, while green squares indicate additional high-$z$ galaxies at low metallicity, for which the metallicity has been measured directly \citep{Willott2025,Cullen2025,Mowla2024}, and the purple pentagons show the median values inferred by the AURORA survey at high-$z$ \citep{Sanders2025Aurora}.
The \cite{Sanders2024} calibration provides a metallicity $4.7\times 10^{-3}~Z_\odot$ for the central region, and an upper limit of $3.9\times 10^{-3}~Z_\odot$ for the $\sim$200~pc extended region (based on the 3$\sigma$ upper limit on [OIII]/H$\beta$).
There is potentially a highly super-solar solution (high metallicity branch) that, in addition to being highly implausible for such a small system at such high redshift, is ruled out (shaded region) by the weakness of the low ionization lines ([OII]3727 and [SII]6720). This is discussed more in detail in  Appendix \ref{app:high-Z}.

We note that we are adopting an extrapolation of the \cite{Sanders2024} calibration to low metallicities, just as other studies of other similar extremely low metallicity systems.  However, unlike other works \citep{Nakajima2025_lowZ,Morishita2025_lowZ}, the calibration adopted by us is more conservative. Indeed, other calibrations used by other works, would give even lower metallicities. As an example, the photoniozation models by \cite{Nakajima2022} with the highest ionization parameter ($log(U)=-0.5$, green long dashed line in Fig.~\ref{fig:met}a), used by \cite{Nakajima2025_lowZ} and \cite{Morishita2025_lowZ} in their low metallicity galaxies at high-$z$, would give an even lower metallicity, by  0.2--0.3~dex.

Fig.~\ref{fig:met}b shows also the calibration using the parameter 
$\hat{R}=0.46~\log{([OII]/H\beta)}+0.88~\log{([OIII]/H\beta)}$, initially proposed by \cite{Laseter2024metallicitycalib}, which has the advantage of being more steeply declining at low metallicities. In this case we only have upper limits on this parameter (as [OII]3727 is not detected),
but which give upper limits on the metallicity ($Z<10^{-2}~Z_\odot$) consistent with the values in Fig.~\ref{fig:met}a. Also in this case we are using a conservative calibration obtained at high redshift - using the alternative calibration by \cite{Scholte2025} (red dashed line in Fig.~\ref{fig:met}b) would give upper limits that are 0.3--0.4 dex lower.

The inferred constraints on the metallicity are summarized in Table \ref{tab:properties}, together with other relevant properties for QSO1.

\begin{table}
    \begin{tabular}{l}
    \hline
    $\log{(Z/Z_\odot)} ={ -2.32^{+0.09}_{-0.12}} ~~~[R<150~pc]$\\
    $\log{(Z/Z_\odot)} < { -2.41} ~~~[150~pc<R<300~pc]$\\
    $^a A_V <0.5$\\
     $^b~log{(M_{BH}/M_\odot)} = 7.7\pm 0.3$ \\
     $^b~L/L_\mathrm{Edd}\approx 0.03$\\
     $^b~M_{star}<2\times 10^7~M_\odot$\\
     $^b~M_{BH}/M_{star}>2$\\
     \hline
    \end{tabular}
\caption{Physical properties inferred for QSO1.
    Notes:  $^a$~From \citet{DEugenio2025};
    $^b$~From \citet{Juodzbalis2025_directBH}.
    }
    \label{tab:properties}
\end{table}


\section{Dust extinction}
\label{sec:dust}

The ratio between the narrow components of H$\alpha$ and H$\beta$ has been measured by \cite{DEugenio2025}
to be 
$2.6^{+0.6}_{-0.5}$, i.e. 
consistent with Case B, hence indicating no or negligible dust extinction ($A_V<0.5$). This is fully consistent with our finding of extremely low metallicity, which is expected to  be accompanied by little dust.

Within the context of this section it is worth mentioning that a very low metallicity is also independently derived  by \cite{DEugenio2025}, from the properties of the absorbing gas, who infer a very low dust content, specifically $Z\xi_d < 3\times 10^{-4}$, where $\xi_d$ is the dust-to-metal ratio. Assuming an evolution of $\xi_d$ as a function of $Z$ as obtained by \cite{Konstantopoulou2024}, this would imply $Z< 4\times 10^{-3}~Z_\odot$, consistent with the upper limit that we have inferred in the outer region of QSO1 from the [OIII]/$\beta$ ratio.

\section{Outflow and black hole mass}
\label{sec:BH_mass}

We will see that our inferred metallicity measurement already provides important constraints on models and simulations. However, these constraints become even more stringent if accompanied by constraints on the black hole mass. Therefore, in this section, we provide a brief overview of the  BH mass measurement presented for QSO1 and also take the opportunity to discuss more the profile of the broad component of the Balmer lines, although an in depth discussion of these aspects is beyond the scope of this paper.

The black hole mass in QSO1 was first estimated by \cite{Furtak2023_AGN} to be $\log{(M_{BH}/M_\odot)}\approx 7.5$ based on the profile of H$\beta$ in their  low resolution prism spectrum and assuming the local virial relations \citep{Reines2016}.
These relations allow the black hole mass to be inferred based on the broad Balmer emission lines' luminosity and their width or dispersion.
However, in their low resolution spectrum the broad component of H$\beta$ was marginally resolved and totally blended with the narrow component.

\cite{Ji2025} derived a BH mass in the range $\log{(M_{BH}/M_\odot)}=7.2-7.6$, depending on the assumed extinction and methodology, and based on the broad component of H$\beta$ measured in their high resolution spectrum, fitted with a single broad Gaussian. 

As we discussed, fitting the broad H$\beta$ with a single Gaussian gives the residual of an intermediate component, slightly blueshifted relative to the narrow component. The fact that this component is spatially resolved on scales of a few 100~pc strongly suggests that this component is tracing an outflow.
\cite{Juodzbalis2025_directBH} investigated the higher S/N and higher resolution spectrum of H$\alpha$; they also identify the intermediate component of H$\alpha$ and they also find that it is extended on scales of a few 100~pc (although the angular resolution at the wavelength of H$\alpha$ is 50\% lower than at the wavelength of H$\beta$).

This intermediate component is not seen in [OIII], but it would not be detectable if it had the same ratio to H$\beta$ as the narrow components. This suggests that also the outflow is very metal poor. This is a very fascinating aspect, as it may highlight  early feedback processes in these primeval black holes. However, a more extensive analysis of the outflow in this source is beyond the scope of the paper and will be presented in a separate work. Here we only mention that if this intermediate component was included in the computation of the black hole mass it would have reduced it, as the effective width of the combined broad line would have been smaller. 

By taking only the spatially unresolved, broader component of H$\alpha$ (i.e. the one most likely associated with the BLR),
and correcting for the mild extinction inferred from the narrow Balmer decrement, \cite{DEugenio2025} estimate a black hole mass of
$\log{(M_{BH}/M_\odot)}=7.1 \pm 0.3$, i.e.  consistent with the estimate from H$\beta$.

All these estimates are based on single epoch measurements, relying on virial relations that are calibrated locally, either through reverberation mapping or direct measurements \citep[e.g.][]{Reines2016,DallaBonta2025}. There has recently been much debate about whether such relations are adequate at high reshift especially for LRDs, with works claiming that such relations may overestimate the BH masses by two orders of magnitude \citep{Rusakov2025,Naidu2025,Chang2025absorpt}, while others claiming that the BH masses might be underestimated by large factors when accreting close or beyond the Eddington limit \citep{Marconi2008,Marconi2009}. The direct black hole mass determination of a luminous quasar at $z=2.3$, through the spatially resolved observation of its BLR with interferometry, has found a value consistent with the virial relations within a factor of 2.5, when using H$\alpha$. However, LRDs, and more generally AGN at even higher redshifts, may behave differently. 

Yet, specifically for QSO1, \cite{Juodzbalis2025_directBH} used the BlackTHUNDER data to resolve the rotation curve traced by the narrow lines in the inner region, down to the central few tens pc, also thanks to the gravitational magnification. They leverage both spectro-astrometry and detailed dynamical modelling using the MOKA3D framework \citep{Marconcini2023}, which reproduces both the velocity field and the light distribution by taking into account also the PSF smearing. They find that the kinematics is consistent with pure Keplerian rotation around a point mass of 
$\log{(M_{BH}/M_\odot)}=7.7 \pm 0.3$. This is consistent within the uncertainties, with the black hole masses estimated via virial relations.

As already mentioned in the introduction, \cite{Juodzbalis2025_directBH} also noted that the Keplerian curve leaves little room for any stellar component in the host galaxy, yielding a tight dynamical upper limit on the stellar mass of $2\times 10^7~M_\odot$. This results into an unprecedented lower limit on the black hole to stellar mass ratio of $\rm M_{BH}/M_{star}>2$.

The upper limit on the stellar mass given above is assuming an exponential stellar disc (Sersic index = 1); the upper limit becomes even lower  if one considers the scenario of a very compact nuclear star cluster \citep[see ][]{Juodzbalis2025_directBH}.

In the rest of this paper we will adopt the direct measurement of the BH mass obtained by \cite{Juodzbalis2025_directBH} and, conservatively, the black hole to stellar mass lower limit inferred by them when assuming an exponential disc.


\begin{figure*}
	\centering
	\includegraphics[width=0.9\linewidth]{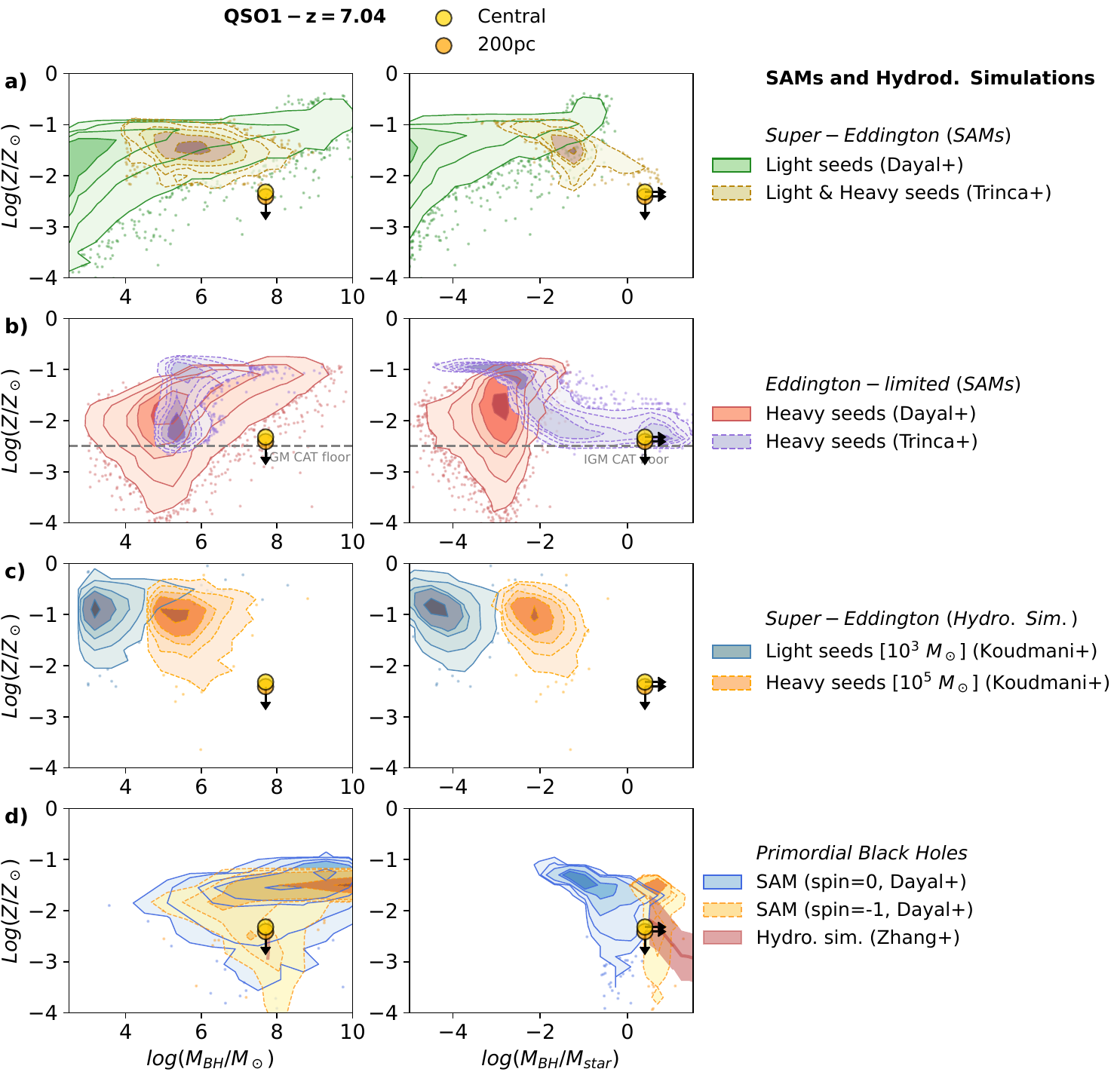}
	\caption{Comparison of the QSO1 properties with Semi-Analytical Models (SAMs) and Hydrodynamical Simulations on the metallicity versus $M_{BH}$ (left) and metallicity versus $M_{BH}/M_{star}$ (right) diagrams. Description of all models are provided in the text and in Appendix~\ref{app:models}. Contours enclose 1\%, 30\%, 68\%, 84\%, 95\%, and 99\% of the models or simulations. Models and simulations outside the 99\% countours are plotted individually. For
	the PBH hydrodynamical simulations, the red line shows the evolution of a representative case, and the shaded regions indicate the uncertainty on the metallicity. The golden and orange circles show the properties of QSO1 in the central aperture and in the outer annulus ($\sim 200~\rm{pc}$ from the centre), respectively (upper limits are at 3$\sigma$).}
	\label{fig:models}
\end{figure*}

\section{Comparison with models and simulations}
\label{sec:models}

In most scenarios of black hole-galaxy co-evolution it is difficult to grow a black hole with masses in excess of $10^7~M_\odot$ while keeping the metallicity of the interstellar medium as low as $10^{-2}~Z_\odot$. Therefore, the findings presented in this paper are indicative of some specific route that allowed the black hole to form and grow in a very un-evolved stellar system. 

In this section we compare our results with predictions of various models and simulations, which are described more in detail in Appendix \ref{app:models}, and which are not meant to be exhaustive.

Fig.~\ref{fig:models} shows the location of QSO1 on the metallicity versus black hole mass plane, and versus black hole to stellar mass ratio. The large symbols show the locations of QSO1 as inferred by our analysis (Tab.\ref{tab:properties}), both considering the metallicity measured centrally (golden) and in the 200~pc annulus (orange), together with the BH mass measurement and BH to stellar mass ratio constraint directly inferred by \cite{Juodzbalis2025_directBH}, as discussed more in detail in Sect.\ref{sec:BH_mass}.
In terms of $M_{BH}/M_{star}$ ratio it is not totally obvious whether we should take the total upper limit on $M_{star}$ or values estimated matching the extraction apertures. Different models and simulations generally provide integrated values and generally do not resolve these small scales; additionally, as discussed, the upper limit on the stellar mass depends on the assumed profile (exponential disc vs. nuclear star cluster).
For the sake of simplicity, we take the global upper limit inferred by \cite{Juodzbalis2025_directBH} on the stellar mass assuming an exponential profile. We recall that this upper limit was obtained by modelling the rotation curve traced by the narrow H$\alpha$, which gave an upper limit on the dynamical mass of the host galaxy of $M_{dyn}<2\times 10^7 M_\odot$, and therefore a tight upper limit on the stellar mass. This gives a tight lower limit on the black hole to stellar mass ratio of $M_{BH}/M_{star}>2$, that, for simplicity, we associate to both the central region and the 200~pc annulus,
making it clear that this is a conservative lower limit.

In Fig.~\ref{fig:models}a the contours show the distribution of semi-analytical models (SAMs) at $z$=7 which assume super-Eddington accretion. In all plots the outermost contour includes 99\% of the distribution.
 Green solid contours are for models from the DELPHI SAMs  \citep{Dayal2024DELPHI} with light seeds ($100~M_\odot$) that are allowed to accrete at up to 5 times higher than the Eddington limit. The dashed brown contours are from the SAM CAT models \citep{Trinca2024} involving both light and heavy seeds ($10^2-10^5~M_\odot$), allowed to accrete up to tens of times above Eddington for short phases.
 These SAM models struggle to reproduce the observations. However, a small fraction of the CAT SAMs can potentially be marginally consistent with the properties of QSO1, altough in less than 1\% of the cases. This tail is part of the models that started with heavy seeds.
 
One should however consider that, as we discuss in the next section, QSO1 is not a rare object at z$>$7 -- it is probably fairly common among AGN at z$>$7. Therefore, successful models and simulations should also reproduce a relatively large fraction of these objects. On the other hand, these SAMs (as well as simulations discussed further below) were optimized to reproduce populations of black holes that did not include cases like QSO1. Probably, future optimization in the assumptions of these SAMs will be more successful in reproducing the observational properties of QSO1.

Fig.~\ref{fig:models}b shows the DELPHI and CAT models at $z$=7 in the scenario of heavy seeds ($10^3-10^5~M_\odot$) whose accretion is Eddington limited. These scenarios can reach lower metallicity at $z$$\sim$7 relative to the super-Eddington model. The primary reason is that in the super-Eddington models the excess of gas inflow required to boost BH accretion also results in boosting star formation that rapidly enriches the ISM.
One should note that the CAT SAMs have a metallicity floor that is given by the modelling assumption that the IGM metallicity at this epoch is $\log{(Z_{IGM}/Z_\odot)}=-2.5$ (horizontal dashed line).
Both these Eddington-limited models fail to either achieve the required black hole mass, or the required $M_{BH}/M_{star}$ ratio at the observed metallicity. Essentially these models do not manage to grow the black hole large enough, unless they are accompanied by significant star formation in the host, which both significantly increases the metallicity and significantly reduces the black hole to stellar mass ratio.

Fig.~\ref{fig:models}c illustrates the case of the \textsc{aesopica} hydrodynamical simulations based on the \textsc{fable} model \citep{Henden2018} with modifications to the BH modelling \cite{Koudmani2022} to allow for very high accretion rates up to 10 times the Eddington limit. Orange dashed contours show the case of heavy seeds ($10^5~M_\odot$), while gray solid contours show the case of the upper boundary for light seeds ($10^3~M_\odot$). In these hydrodynamical simulations light seeds, even accreting at super-Eddington rates, are completely excluded. The case of heavy seeds also struggles to reproduce QSO1 because of its low metallicity, but also because of the high $M_{BH}$ and high $M_{BH}/M_{star}$. 
However, as already discussed, one should take into account that these simulations, as well as the SAMs discussed above, are developed with assumptions aimed at reproducing black holes with properties quite different than QSO1. Future upgrades may provide a better match with QSO1. For instance, heavier seeds, with masses up to $\sim 10^6~M_\odot$ (still consistent with, for instance, the Direct Collapse scenario), may result in a distribution closer to QSO1. Additional gas inflows, kept warm by the UV radiation and prevented from forming additional stars, could help to significantly dilute  the metallicity to the level observed in QSO1.

Fig.~\ref{fig:models}d shows the case of ``primordial black holes'' (PBHs), i.e. black holes that, according to some scenarios, might have formed in the ultra-early universe, in the first second after the Big Bang, possibly reaching back to the inflationary era \citep[e.g.][]{Hawking1971,Carr1974,Escriva2024,Riotto2024_PBH,Zhang2025_PBH,Belotsky2019_PBH}.
Although the relevance of PBHs as dominant dark matter candidates has been mostly ruled out \citep[e.g.][]{Ziparo2022_PBHconstraints,Casanueva2025_PBHconstraints}, they are still considered viable as seeds for the subsequent growth of massive black holes.
In this scenario, black holes would be the very first structures formed in the Universe, much earlier than stars and galaxies.
The expected mass function of primordial BHs is very uncertain, also because they are expected to be very clustered \citep{Belotsky2019_PBH,Zhang2025_PBH}, possibly causing them to rapidly merge hierarchically and, therefore,  shift their distribution towards higher masses soon after the Big Bang. The blue, solid contours in Fig.~\ref{fig:models}d show a version of the {\sc phanes} analytic model \citep{dayal2024_pbh,Dayal_Maiolino2025} in which the growth of halos and their baryonic components is seeded by primordial BHs, and in the sub-case of non-spinning black holes (Appendix \ref{app:models-delphi}).  This model can reproduce QSO1 within a few percent of the distribution, i.e. reasonably well considering the uncertainties associated with this class of black holes. The case of counter-rotating PBHs (spin $=-1$) in the same model is shown with orange, dashed contours; considering that we have only a lower limit on $M_{BH}/M_{star}>2$, these models can potentially reproduce QSO1 even better. However, one should consider that counter-rotation can be maintained by the black hole only for a few Eddington times.

The red track in Fig.~\ref{fig:models}d
shows the result of hydrodynamical simulations \citep{Zhang2025arXiv250317585Z}, which follow the evolution of primordial black holes, assuming the case of a $5\times 10^7~M_\odot$ PBH. This mass is higher than the canonical $10^6~M_\odot$ scale preferred for PBHs, given by the fundamental-physics connection to the electron-positron annihilation epoch in the ultra-early universe \citep[e.g.][]{Carr2020}; however, as mentioned, PBHs are expected to be highly clustered and may rapidly merge resulting in much heavier seeds \citep[e.g.][]{Zhang2024_BiasedPBH}. The track in Fig.~\ref{fig:models}d represents snapshots between $7<z<9.5$. As discussed in Appendix~\ref{app:models-zhang}, the metallicity in these simulations is determined at the post-processing stage (which is a limitation that will be tackled in a follow-up study). 
As illustrated, in these hydrodynamical simulations the assumed BH mass increases very little in this short epoch, whereas the metallicity and $M_{star}$ evolve rapidly with time. Overall, these hydrodynamical simulations incorporating PBHs can also potentially explain the properties of QSO1, although subject to the specific assumptions underlying the metallicity evolution and the assumed initial mass of the PBH.

Summarizing, the fact that heavy seeds or super-Eddington models/simulations reproduce the properties of QSO1 only in rare cases (or fail completely), may highlight some potential problems in these scenarios. As mentioned, it may indicate that these models require additional, major pristine gas inflows to dilute the metallicity. Additionally, stronger feedback processes may be at work, both ejecting metals more efficiently and/or suppressing star formation (and hence the production of metals) more effectively. Alternatively, the underlying BH seeding and growth assumptions may not be adequate to reproduce this population of objects in the early Universe. Heavier seeds, or seeding at later times may be needed.

We re-iterate that those SAMs and simulations were not optimized to reproduce the properties found at z=7 in QSO1.
Indeed, for instance, some of these models manage to reproduce the properties of the much more luminous quasars at z$\sim$6--7 \citep{Bennett2024} -- although the metallicity of the host galaxies in distant quasars has not been studied in detail yet, based on their spectral properties they are inferred to have metallicities close to solar \citep[e.g.][]{Juarez2009,Wang2022metallicityAGN}, and, while they are overmassive in terms of black hole to stellar mass ratio \citep{Stone2024,Marshall2025}, they are not as extreme as QSO1.
On the contrary, reproducing the extremely low metallicity and the very high black hole to stellar mass ratio in QSO1, and possibly in other AGN at similar redshifts (see next section), will probably require additional development of those models.
Within this context it is interesting that \cite{Jeon2025b} has more recently developed a SAM more specifically tailored to LRDs and also specifically investigating objects like QSO1. While their light seeds scenario (including super-Eddigton accretion) does not manage to reproduce QSO1, their DCBH prescription can reproduce the properties of QSO1 reasonably well; however, their BH mass function is nearly 2 dex below recent observations at high redshift \citep{Geris2025}. Another problem of the DCBH scenarios is that most of them require a nearby source of UV radiation for photodissociating H$_2$, while QSO1 is fairly isolated without strong sources of UV radiation in the vicinity, or post-starburst galaxies that may have produced UV radiation in the past. An additional problem of DCBH models specifically for QSO1, as pointed out by  \cite{Juodzbalis2025_directBH}, is that their early growth is limited by the baryon fraction in an atomically-cooling halo 
\citep{BrommYoshida2011_Review,Pacucci2015_DCBH};
this sets an upper limit to the black hole to dynamical mass ratio of $\sim 0.1$, which is more than 1~dex lower than observed in QSO1, as already discussed earlier on \citep{Juodzbalis2025_directBH}.

On the other hand, the fact that
scenarios invoking PBHs can explain reasonably well the properties of QSO1-like system, including their seemingly high space density discussed in the next section, indicates that the physics behind this model may possibly account for a significant population of early BHs. As already discussed, the large mass in QSO1 might be problematic to achieve with pure PBHs without invoking significant accretion, which may have also resulted in star formation and chemical enrichment; however, rapid merging of PBHs (which are highly clustered) can potentially increase their mass by a large factor without resulting in star formation and chemical enrichment.

An intermediate scenario which could potentially account for the properties of QSO1 is that of the ``Not Quite Primordial Black Holes'' \citep[NQPBHs;][]{Qin2025_NQPH}. These are putative massive seeds formed at $z>200$, resulting from the direct collapse of clouds, as a consequence of the CMB at that redshift being energetic enough to photodissociate H$_2$. This scenario may help to achieve a higher mass, but even in this case accretion or merging may be required to attain the BH mass of QSO1 for a significant fraction of the objects, as discussed in the next section. More generally, this scenario requires further investigation to explore and predict more quantitatively whether the properties of QSO1 can be reproduced or not.

Yet another intriguing possibility is that QSO1 could be a BH ejected from a nucleus where there were 3 black holes that interacted and ejected the least massive one. Since this scenario was initially conceived by \cite{Saslaw1974}, several groups have further developed this idea and predicting that this population might be relatively abundant \citep[e.g.][]{Volonteri2003}. However, it is not clear how much gas from the parent galaxy may remain bound to the ejected black hole, or how much gas may accrete from the intergalactic medium. Leaving aside these aspects, it is unlikely that a relatively powerful AGN aligned with the lensing cluster is one of this special cases of ejected black holes. However, certainly this scenario deserves further exploration, both theoretically and observationally.

Finally, we caution that  all these comparisons with models are done assuming that, at a given BH mass, the accretion rate is not dependent on metallicity. Additionally, we have been comparing our results for an accreting AGN \citep[although at a fairly low rate of $L\L_{Edd}\sim0.03$, according to ][]{Juodzbalis2025_directBH}
with models and simulations of the whole population of black holes -- AGN duty cycles certainly play a role in their detectability, which should be taken into account in future studies.

Regardless of the specific model and simulation, the finding that QSO1 is embedded in such an extremely low metallicity gas, and that it is nearly ``naked'', with little (if any) stellar mass around it ($M_{BH}/M_{star}>2$), suggests that it is a massive seed in the earliest phase of accretion.


\section{How common are extremely metal poor AGN and LRDs in the early Universe?}
\label{sec:density}

It is too early to quantitatively assess the fraction and density of extremely metal poor AGN, and in particular LRDs, in the early Universe, and certainly more observations and statistics are needed. 
However, we note that QSO1-like BHs cannot be so uncommon. Indeed, QSO1 is one of the two BHs found at $z$$>$7 in the whole UNCOVER survey \citep{Greene2024}, and the only one with $M_{BH}<10^8~M_\odot$.
Additionally, it is found in a very small sky area close to the caustic behind the Abell 2744 field (within $\sim$30$''$) that could produce multiple images, meaning that the volume density of these type of objects cannot be low.

Although the number and quality of observations of low luminosity AGN with intermediate black hole masses at z$>$7 are still inadequate to draw conclusions, it is important to compare QSO1 with the few other AGN discovered at z$>$7. Aside from QSO1, there are six type 1 AGN found at $z$$>$7 \citep{Greene2024,Kokorev2023_AGN,Maiolino24_GN-z11,Tripodi2024_lowZ,Ubler24,Taylor2025LRDz8,Naidu2025}.
One of them has an estimated very low metallicity of $Z<0.1~Z_\odot$ \citep{Tripodi2024_lowZ} and two more have very weak [OIII]5007
(although their [OIII]/H$\beta_N$ is difficult to constrain due to low spectral resolution,
\citealt{Taylor2025_capersLRD,Naidu2025}), so these might share similar properties as QSO1. Therefore, although statistics are still low, about half of the AGN discovered by JWST at z$>$7 seem to have very low metallicities, possibly as extreme as QSO1.

Summarizing, it is very unlikely that QSO1 is an outlier, and implausible that it is in the tail of a distribution with probability to be found of less than a few percent. QSO1-like systems
must be fairly common among intermediate mass BHs ($M_{BH}$ a few times $10^7~M_\odot$) at $z$$>$7.

However, at the same time, one should also note that the other three AGN at z$>$7 have [OIII]/H$\beta$ significantly higher than in QSO1, and likely have higher metallicity.  
Therefore, the origin of these other BHs may well be different from the one in QSO1. Most of them also have masses much higher than QSO1 and may have formed and grown through different channels \citep[e.g.][]{Jeon2025}. These three AGN at z$>$7 with high [OIII]/H$\beta$ would certainly be easier to reproduce by some of the models presented in the previous section, not only because  of the probably higher metallicity, but also because they do not have black hole to stellar mass ratio as high as QSO1.

\section{Conclusions}
\label{sec:conclusions}

We have analysed the deep, high resolution NIRSpec-IFS spectroscopy of Abell2744-QSO1,  obtained as part of the BlackTHUNDER JWST Large Programme. QSO1 is a strongly lensed Little Red Dot (LRD) at z=7.04, hosting a massive black hole ($M_{BH}\approx 5\times 10^7~M_\odot$, measured directly), and with $M_{BH}/M_{star}>2$. We have focused on the narrow components of the emission lines and, in particular H$\beta$ and [OIII]. The high resolution spectroscopy of the new data allows the clear disentangling of the narrow and broad components of H$\beta$. Additionally, the high sensitivity of the data, together with the integral field mode and the strong lensing shear, allow the spatial extent of H$\beta$ to be traced.
The main results of our analysis are the following:

\begin{itemize}

    \item While the continuum is spatially unresolved, the narrow component of H$\beta$ is spatially extended out to $\sim 400$~pc.

    \item In the central region ($R<150~{\rm pc}$) the [OIII]5007 line is extremely weak relative to the narrow component of H$\beta$, with a ratio $[OIII]5007/H\beta=0.66$, one of the lowest ever observed in the ISM of distant galaxies. In an annulus with $150 {\rm \,pc}<R<300 {\rm \,pc}$ H$\beta$ narrow is still clearly detected while [OIII]5007 is not significantly detected, resulting $[OIII]5007/H\beta<0.41$.
    
    \item As in other similar galaxies at high-$z$, the simplest interpretation of the low $[OIII]5007/H\beta$ ratio is very low metallicity of the gas in the vicinity of the black hole (while we show that density or ionization effects cannot play a significant role). We estimate $Z\approx 4.7\times 10^{-3} ~Z_\odot$ in the central region ($R<150~{\rm pc}$), and $Z<3.9\times 10^{-3}~Z_\odot$ in the annulus at about 200~pc.

    \item We have discussed that massive black holes with such low metallicity are probably fairly common in the early universe, although more data are needed to properly assess their occurrence.

    \item It is challenging for most models to account for such a chemically unevolved system that host a black hole that is already so massive.
    More specifically, we have shown that most models and simulations struggle to reproduce the properties of QSO1, although some of them can reproduce the observed low metallicity, together with the high $M_{BH}$ and high $M_{BH}/M_{star}$. This may indicate that some revision in the assumptions of these models may be required (e.g. additional pristine inflow diluting the metallicity, more massive seeds, or stronger feedback). Primordial black holes scenarios seem to better reproduce the properties of QSO1, but even these scenarios require further developments to self-consistently  explore the black hole and metallicity evolution. The scenarios of Not Quite Primordial Black holes, or a black hole ejected from the interaction with two other merging black holes, should also be explored more thoroughly.

\end{itemize}

More qualitatively, the finding of extremely low metallicity, together with the high $M_{BH}$ and extremely high $M_{BH}/M_{star}$, indicates that this LRD, and probably many others at similar redshifts, are tracing the earliest accretion phase on massive black holes seeds \citep{inayoshi2025}.
There is clearly the need of additional efforts in modelling and simulating the various black hole scenarios in order to properly, and self-consistently reproduce the properties of QSO1 and similar LRDs and AGN in the early Universe.

While we have discussed that other AGN at high redshift seem to be characterized a very low metallicity similar to that observed in QSO1, these other cases need additional observations to be confirmed. Yet, other AGN discovered by JWST at $z$$>$7 are characterized by much higher metallicity relative to QSO1, and these may have formed from different seeds and evolved through different routes relative to QSO1.

\section*{Acknowledgements}

This paper is dedicated to Avishai Dekel, an innovative and inspirational thinker, who made seminal contributions to astrophysics.
We thank the anonymous referee for their comments that helped to improve the manuscript.
We thank Andrea Ferrara for useful suggestions. This work is based on observations made with the National Aeronautics and Space Administration (NASA)/European Space Agency (ESA)/Canadian Space Agency (CSA) JWST. The data were obtained from the Mikulski Archive for Space Telescopes at the STScI, which is operated by the Association of Universities for Research in Astronomy, Inc., under NASA contract NAS 5-03127 for JWST. These observations are associated with programme PID 5015. RM, FD, JS, IJ, GJ acknowledge support from the Science and Technology Facilities Council (STFC), by the European Research Council (ERC) through Advanced Grant 695671 ``QUENCH'', by the UK Research and Innovation (UKRI) Frontier Research grant RISEandFALL. RM also acknowledges support from a Royal Society Research Professorship grant. 
SZ, VB and BL acknowledge the Texas Advanced Computing Center (TACC) for providing HPC resources under allocation AST23026. 
GV acknowledges support by European Union’s HE ERC Starting Grant No. 101040227 - WINGS.
H\"U acknowledges funding by the European Union (ERC APEX, 101164796). Views and opinions expressed are however those of the authors only and do not necessarily reflect those of the European Union or the European Research Council Executive Agency. Neither the European Union nor the granting authority can be held responsible for them.
K.I. acknowledges support from the National Natural Science Foundation of China (12233001),  the National Key R\&D Program of China (2022YFF0503401), and the China Manned Space Program (CMS-CSST-2025-A09).

\section*{Data availability}

The NIRSpec data used in this study are publicly available at the STScI MAST archive: \url{https://mast.stsci.edu/portal/Mashup/Clients/Mast/Portal.html}, under GO programme 5015, observation 22.




\bibliographystyle{mnras}
\bibliography{AGN} 


\appendix

\section{Ruling out the high metallicity solution}
\label{app:high-Z}

\begin{figure}
	\centering
    \includegraphics[width=0.9\linewidth]{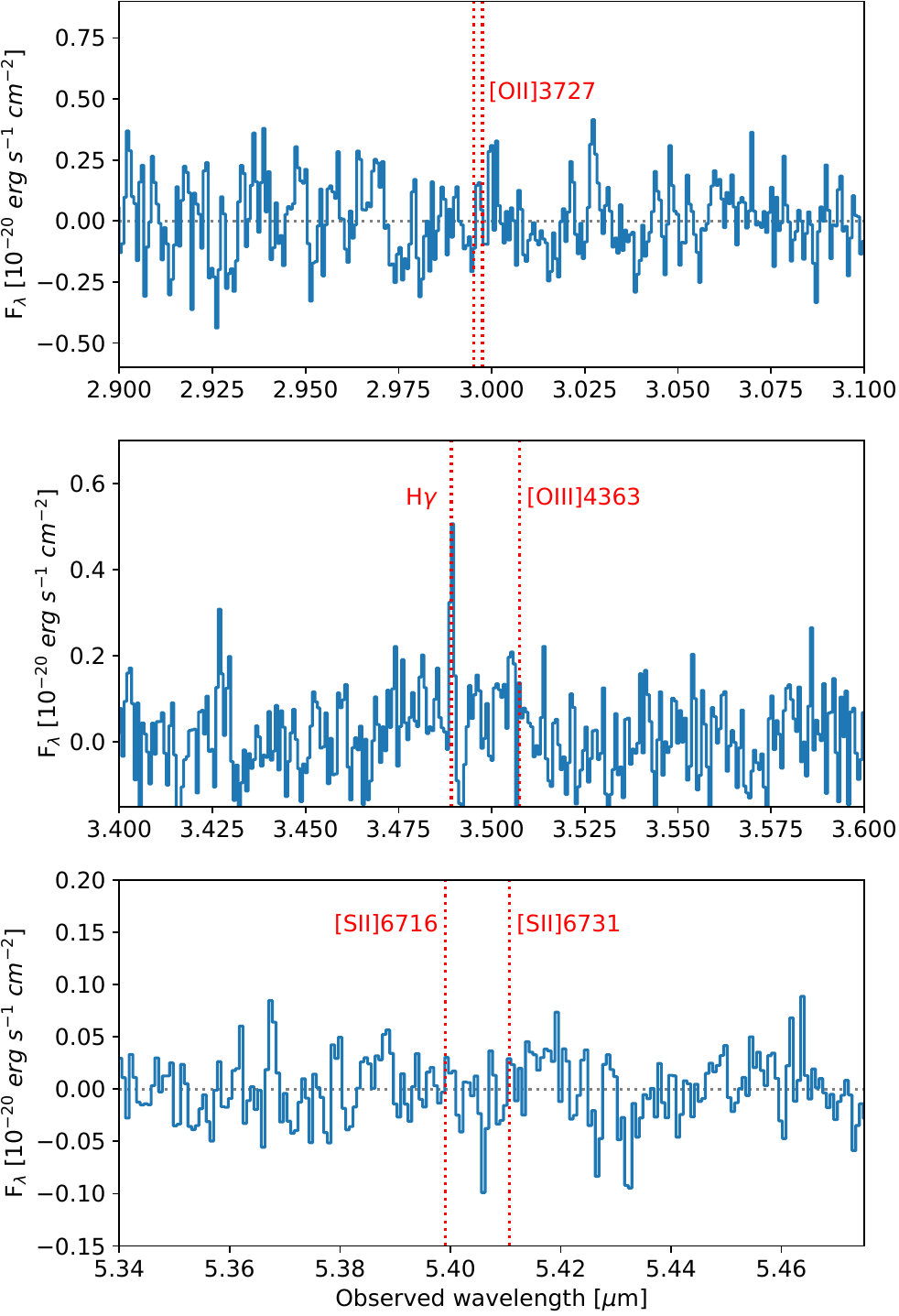}
	\caption{Portion of spectra extracted from the central aperture illustrating the non detections of [OII]3727, [OIII]4363 and [SII]6716,6731.
	}
	\label{fig:oii_sii}
\end{figure}

\begin{figure}
	\centering
    \includegraphics[width=0.9\linewidth]{"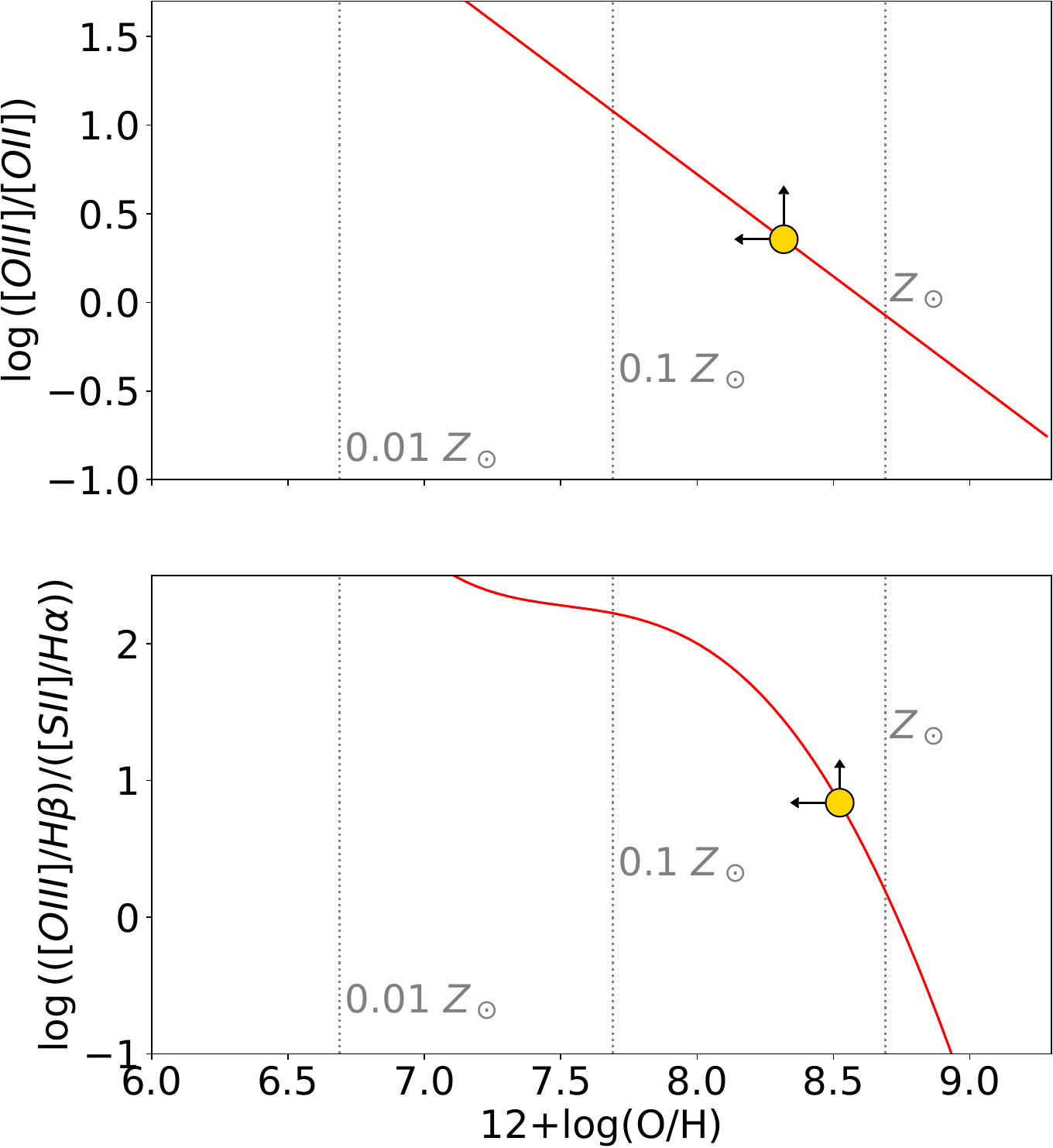"}
	\caption{Constraints on the metallicity from the non-detection of low ionization lines. Top: log([OIII]5007/[OII]3727) lines flux ratio versus metallicity according to the calibration obtained by \citet{Sanders2024} at high-z (red line). Bottom: log(([OIII]5007/H$\beta$)/([SII]6716,6731/H$\alpha$)) as a function of metallicity according to the local relation identified by \citet{Curti2020}. In both panels the golden symbols indicate the $3\sigma$ limits obtained for QSO1, both of which exclude the highly super-solar metallicity solution of the [OIII]/H$\beta$ value.
	}
	\label{fig:met_uplim}
\end{figure}

As discussed in the main text, for a given [OIII]/H$\beta$ ratio the metallicity calibration has two solutions, as illustrated in Fig.\ref{fig:met}. Therefore, in principle, the [OIII]/H$\beta$ value observed in QSO1 could also correspond to extremely high metallicity, more than {\bf three} times solar. Such an extremely high metallicity would be totally implausible for such a low mass system at such high redshift \citep[e.g.][]{Curti2023}. Additionally, the high metallicity solution can be ruled out based on the upper limits on the low ionization lines. Indeed,  as illustrated in Fig.\ref{fig:oii_sii}-top, [OII]3727 is not detected.
The ratio [OIII]/[OII] has a monotonic dependence on metallicity, which has also been calibrated for high-z galaxies \citep{Sanders2024,
Cataldi2025Marta}, as shown in Fig.\ref{fig:met_uplim}-top. The lower limit on the [OIII]/[OII] ratio in QSO1 implies a metallicity $ <0.4~Z_\odot$, hence completely ruling out the supersolar solution of [OIII]/H$\beta$. 
The [SII]6716,6731 doublet is also undetected (Fig.\ref{fig:oii_sii}), and the
 upper limit on [SII]/H$\alpha$ is also informative. Indeed, \cite{Curti2020} showed that the ratio $O3S2 = \frac{[OIII]/H\beta}{[SII]/H\alpha}$ is monotonically decreasing with metallicity. Unfortunately, this diagnostic has not been calibrated at high-z. However, using the local relation, shown in Fig.\ref{fig:met_uplim}-bottom, and the $3\sigma$ lower limits obtained for QSO1 reported in Tab.\ref{tab:fluxes}, we obtain an upper limit on the metallicity of $<0.7~Z_\odot$, once again excluding the highly supersolar metallicity solution of [OIII]/H$\beta$.

\section{The case of AGN excitation of the narrow lines}
\label{app:NLR}

\begin{figure}
	\centering
    \includegraphics[width=0.9\linewidth]{"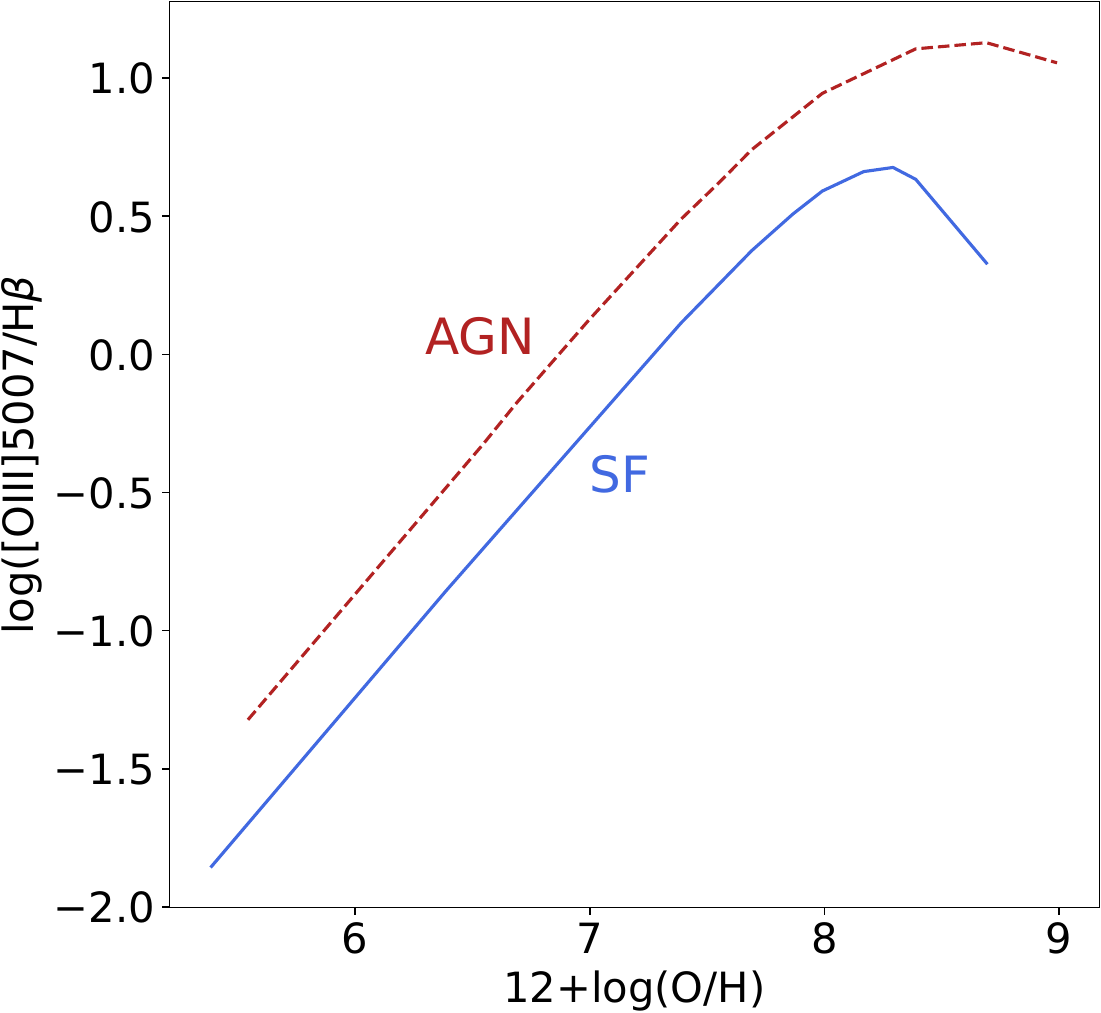"}
	\caption{[OIII]5007/H$\beta$ flux ratio as a function of metallicity for AGN (red dashed) and Star Formation (blue solid) models, as discussed in the text. In the low metallicity (sub-solar) branch, a given [OIII]/H$\beta$ ratio resulting AGN photoionization would result into a metallicity even lower than inferred for the star formation case. 
	}
	\label{fig:cal_AGN_SF}
\end{figure}

In the main text we have assumed, following \cite{DEugenio2025}, that the narrow emission lines are excited by star formation. \cite{DEugenio2025} motivated this by the extremely narrow width of the lines. Here we discuss the scenario in which the narrow lines are excited by the AGN, i.e. are part of a Narrow Line Region, although in this case one would expect broader lines because of feedback from AGN radiation pressure. 

The calibration obtained by \cite{Cataldi2025Marta} for high-z star forming galaxies is not expected to change significantly in the case of AGN. Indeed, high-z galaxies are characterized by very high ionization parameters, similar to the AGN, and tend to have harder ionizing spectrum than local galaxies, akin AGN, because of their reduced iron content \citep{Cameron2024,Strom2017Mosfire}. At the same time, high-z AGN are characterized by softer ionizing spectra than local AGN, akin star forming galaxies, as indicated by the weakness of high-ionization lines \citep{Lambrides2024}. Therefore, the physical conditions and ionization of high-z star forming galaxies and the NLR of high-z AGN are expected to be similar. Indeed, it is known that in many traditional narrow line diagnostic diagrams, high-z AGN and high-z SF galaxies largely overlap \citep{Ubler2023,Maiolino_AGN,Juodzbalis2025type1JADES}; this makes the identification of narrow line (type 2) AGN extremely difficult at high-z \citep{Scholtz2025_type2,
Mazzolari2024b} and indeed the high-z sample used by \cite {Cataldi2025Marta} (and compilation in there) for the calibrations, may include a few high-z AGN according to some diagnostics. Therefore, the low metallicity inferred from the \cite{Cataldi2025Marta} calibration may likely hold also in the case of AGN photoionization of the narrow lines.

In any case, the harder ionization spectrum from an AGN would make the [OIII]/H$\beta$ ratio even higher at a given metallicity, relative to star forming galaxies. We illustrate this in Fig.\ref{fig:cal_AGN_SF}, where we show the example of the expected [OIII]/H$\beta$ ratio in the case of an AGN ionizing spectrum (red dashed) and a star formation ionizing spectrum (blue solid), for a given ionization parameter $log~U = -2$. We use the models presented in \cite{Nakajima2022}, where for the AGN we take the case of an ionizing spectrum modelled with a black body with temperature $2\times 10^5~K$ and power law $\alpha=-2$, while the BPASS models with age 1Myr age  used for star forming galaxies. Clearly, in the low-metallicity branch, in case of AGN photoionization a given [OIII]/H$\beta$ ratio would give an even lower metallicity than inferred by assuming photoionization by star formation.


\section{Ruling out high density scenario for the weakness of [OIII]}
\label{app:high_density}

\begin{figure}
	\centering
\includegraphics[width=\linewidth]{"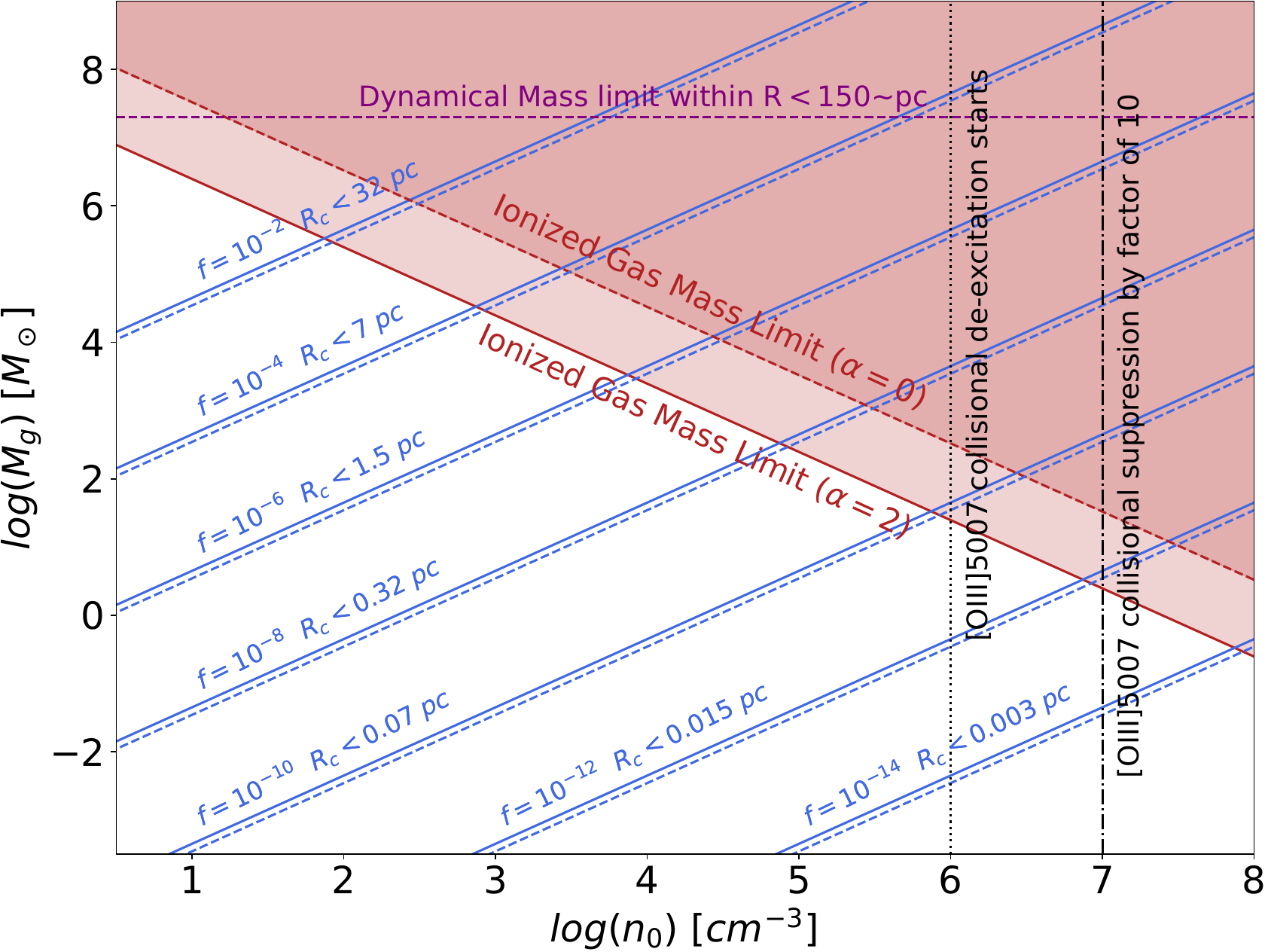"}
	\caption{Constraints on the gas density for the gas emitting the narrow emission lines. The blue lines indicate the gas mass enclosed within the central aperture for as a function of gas density $n_0$ at $R=100pc$, for a power-law radial distribution $n\propto r^{-\alpha}$ with index $\alpha=2$ (solid) and $\alpha=0$ (dashed), for different filling factors (and implied upper limits on the clouds size) as labelled. The red lines indicate the mass of ionized gas constrained by the H$\beta_N$ luminosity for the two power indices. The vertical dotted line indicates the density where collisional de-excitation of [OIII]5007 starts to affect its flux by $\sim 50\%$, the dot-dashed line it the density required to suppress the flux by a factor of 10. Explaining the weakness of [OIII]5007 via collisional de-excitation would require extremely low filling factors. The horizontal purple line indicates the upper limit given by the dynamical mass inferred by \citet{Juodzbalis2025_directBH}.
	}
	\label{fig:density}
\end{figure}

The lack of a broad component of [OIII]5007 is expected, as the gas density in the BLR ($n\sim 10^9-10^{12}~\rm{cm}^{-3}$) is much higher than the critical density of this transition ($n_c = 7\times 10^5~\rm{cm}^{-3}$), hence this line is collisionally de-excited in the BLR, relative to H$\beta$. The weakness of [OIII]5007 relative to the {\it narrow} component of H$\beta$ is instead much more difficult to explain, and actually unphysical, in a high density scenario, as discussed in the following.

 Although the densities of the ionized ISM are found, on average, to increase in high-z galaxies, they are still found to be below $<10^5~\rm{cm}^{-3}$ \citep{Isobe2023density,Topping2025} (one case found to have $n>10^6~\rm{cm}^{-3}$ has actually been ascribed to the BLR of an AGN, with $n\sim 10^{10}~\rm{cm}^{-3}$ \citealt{Maiolino24_GN-z11}). In any case, we can directly provide constraints on the density of the gas in the ionized gas of QSO1 that produces the narrow lines. We focus on the spectrum extracted from the central aperture ($R<150~\rm{pc})$; the arguments would be even stronger for the larger aperture. We assume a radial density distribution following a powerlaw

$$
n(r) = n_0~\left( \frac{r}{r_0}\right) ^{-\alpha}
$$

where $n_0$ is the density at a reference radius $r_0$. We take $r_0=100~\rm{pc}$, i.e. well within the aperture. We also assume that the clouds with such a density profile have a filling factor $f$ defined as
$$
f = \frac{V_{clouds}}{V_{tot}} = \frac{N_{clouds}R_c^3}{R_{max}^3}
$$
where $V_{tot}$ is the volume enclosed in the aperture, assumed spherical with radius $R_{max}$, $V_{clouds}$ is the volume occupied by the ionized clouds, and $N_{clouds}$ is the number of clouds and $R_c$ is their radius.
The mass of gas contained within the aperture is given (assuming a medium made only of hydrogen for simplicity)
$$
M_g = \int _{R_{min}}^{R_{max}}n(r)~m_p~f~4\pi r^2~dr
$$
where $m_p$ is the proton mass.
For $R_{min}$ we take 5~pc, as smaller radii would make the line broader because they would be completely within the sphere of influence of the black hole; we however note that selecting even smaller inner radii would make the arguments even stronger.
The blue lines in Fig.\ref{fig:density} show the implied gas mass as a function of density for different values of the filling factor $f$. We also indicate with labels, for each filling factor, the implied maximum size of the clouds obtained by assuming the extreme case that all gas is contained in a single cloud ($N_{clouds}=1$), while in reality the clouds would be much smaller.
We then derive the mass of ionized gas inferred from the luminosity of $H\beta_N$ 
$$
\frac{M_{ion}}{M_\odot} =
10^9 ~\frac{L(H\beta_N)}{10^{43}erg~s^{-1}}~
\frac{100~cm^{-3}}{n_0}~\frac{3-2\alpha}{3-\alpha}~r_0^{-\alpha}~\frac{R_{max}^{3-\alpha}-R_{min}^{3-\alpha}}{R_{max}^{3-2\alpha}-R_{min}^{3-2\alpha}}
$$
We have assumed a temperature T=10,000~K,  the results would be even tigther if assuming a higher temperature.
The resulting constraint on the mass of ionized gas is shown with a red solid line in Fig.\ref{fig:density}, in the case of $\alpha=2$. We also show the extreme (unlikely) case of $\alpha=0$ (i.e. uniform distribution) with a red dashed line.
In order for the gas to have a density higher than $10^6~\rm{cm}^{-3}$ (vertical black dotted line), at which [OIII]5007 starts to be collisionally suppressed by a factor of 1.5 (which would still not affect significantly our results) the ionized clouds should have an extremely small filling factor ($f<10^{-10}$) and should have sizes much smaller than $0.1~pc$. We have used Cloudy models to infer that, in order to  suppress [OIII]5007 by a factor of ten, which would be needed to entirely ascribe to collisional de-excitation the lower [OIII]/H$\beta$ ratio relative to the other AGN and galaxies at z$\sim$7, the ISM would need to reach densities higher than $10^7~\rm{cm}^{-3}$. As illustrated in Fig.\ref{fig:density} (dot-dashed vertical black line) this would require even more extreme filling factors, even smaller clouds, and a total gas mass of less than a few solar masses.

\begin{figure}
	\centering
\includegraphics[width=\linewidth]{"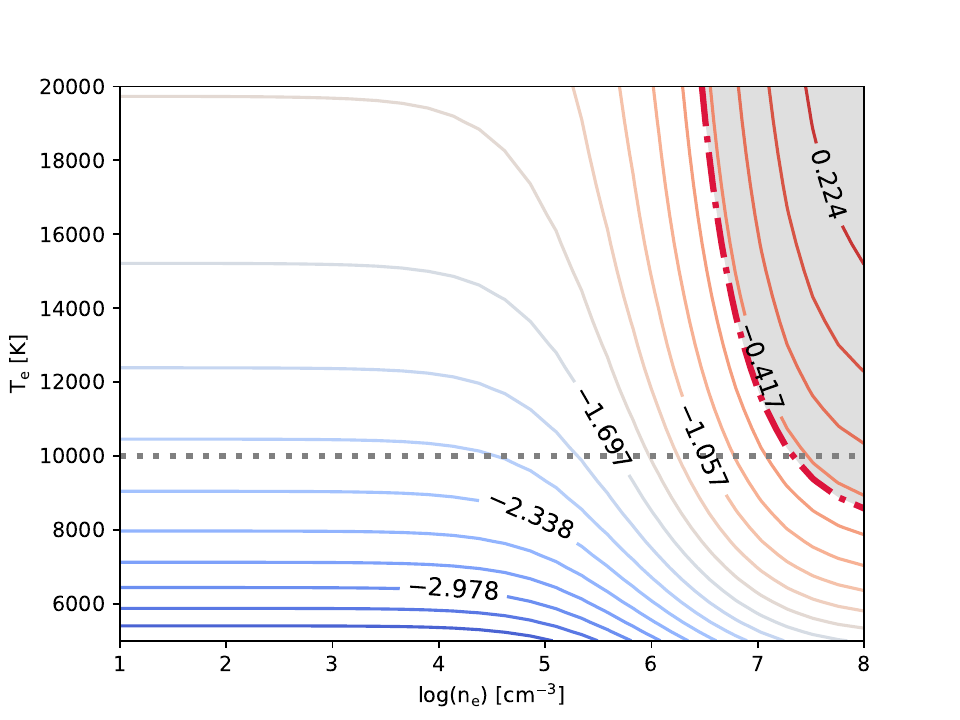"}
	\caption{[OIII]4363/[OIII]5007 flux ratio as a function of temperature and density. The observed upper limit [OIII]4363/[OIII]5007$<$0.33 is marked with a dot-dashed line and indicate densities $<10^7~cm^{-3}$ for reasonable temperatures $T>10^4~K$.
	}
	\label{fig:O3auroral}
\end{figure}

Finally we note that the non detection of [OIII]4363 (as illustrated in Fig.\ref{fig:oii_sii}), which has a critical density much higher than [OIII]5007, also rules out very high densities. Indeed, Fig.\ref{fig:O3auroral} illustrates the variation of  [OIII]4363/[OIII]5007 as a function of temperature and density, estimated with the Pyneb software \citep{PyNeb}. For reasonable temperatures typical of the photoionized gas emitting H$\beta$ ($T>10^4~K$), the 
inferred upper limit in the central region ([OIII]4363/[OIII]5007$<$0.33, Tab.\ref{tab:fluxes}) gives an upper limit on the density of $n<10^7~cm^{-3}$.


\section{Ruling out peculiar ionization scenarios for the [OIII] weakness}
\label{app:low_ion}

In principle, another scenario to explain the weakness of [OIII] relative to H$\beta$, without invoking low metallicity and in the low density regime, is that the gas ionization state is so low that $O^{+2}$ is not produced. This would happen if the ionization parameter $U=Q_i/(4\pi r^2 n c)$ (where $Q_i$ is the rate of ionizing photons) is very low, much lower than typically observed in other galaxies. This would be in contrast with the finding that high-z galaxies are characterized by higher ionization parameters than lower redshift galaxies \citep{Cameron2024}.

Additionally, a low ionization parameter would result in a relatively strong [OII]3727 emission, which is not detected (Appendix \ref{app:high-Z}). We have quantified the latter argument more in detail by using Cloudy photoionization modelling. We have inferred that  (assuming an AGN SED), in the low density regime for the ISM ($n<10^4~\rm{cm}^{-3}$, which gives reasonable filling factors and cloud sizes, Fig.\ref{fig:density}), 
the lower limit on $F([OIII]5007)/F([OII]3727)>3$ (by also taking into accounting an extinction of $A_V=0.66$)  requires an ionization parameter $\log~U > -2.78$.
With such a constraint on the ionization parameter it is not possible to reproduce the observed  low [OIII]/H$\beta$ ratio with the typical metallicity observed in other high-z galaxies of $Z\sim 0.1~Z_\odot$. Reproducing the observed [OIII]/H$\beta$ ratio, while maintaining $\log~U > -2.78$, requires lowering the metallicity, with an upper limit of $log(Z/Z_\odot) < -1.7$, i.e. close to the value inferred from the calibrations.

It is also possible to exclude a low ionization parameter solution based on simple geometrical arguments and based on the gas distribution. The source of UV radiation powering the narrow line (be it AGN or star formation) is unresolved, and confined within $r<30~\rm{pc}$ \citep{Furtak2023_AGN}. For simplicity, and conservatively, we assume it to be point-like. From the H$\alpha$ luminosity (corrected for extinction) we infer a production of ionizing photons $Q_i=1.4\times 10^{54}~\rm{s}^{-1}$. The surface brightness profile within the central aperture is not resolved. However, assuming, as in the previous appendix, a typical powerlaw density profile with power index of $-2$, the luminosity weighted radius within the aperture is 77~pc.
Conservatively assuming an upper limit on the density of
$n<10^4~\rm{cm}^{-3}$
we obtain a lower limit on the ionization parameter of $\log~U>-2.2$. 
With such a high value of the ionization parameter, suppressing [OIII] to the observed value of [OIII]/H$\beta$ would require a metallicity $Z<10^{-1.9}~Z\odot$.

A similar scenario for explaining the low [OIII]/H$\beta$ could be that, without invoking low metallicity,  the ionizing spectrum is deficient in photons energetic enough to ionize O$^+$. However, this scenario would not explain the absence of [OII], as oxygen has about the same ionization potential as hydrogen. Similarly, [NII] and [SII] should be quite strong \citep{Rhea2025}.

The opposite scenario would be that the ionization of the gas is so high that oxygen is mostly in O$^{+3}$. However, it has never been seen, not even in the most extreme AGN, that the ionization of the gas is so high to suppress [OIII] emission \citep{Dors2020}. In any case, such a high level of ionization would result in other high ionization lines, such as CIV and HeII, to be very strong. In particular, using Cloudy modelling, we infer that in order to make [OIII]/H$\beta$ as faint as observed, while keeping the metallicity $lg(Z/Z_\odot)>-1.5$ and assuming an AGN ionizing spectrum, it would require an extremely high ionization parameter, higher than about unity. In such conditions HeII$\lambda$4686 should be about as strong as [OIII], or even stronger, while HeII is undetected.

\section{BPT Diagnostic diagrams}
\label{app:bpt}

\begin{figure*}
	\centering
    \includegraphics[height=0.35\linewidth]{"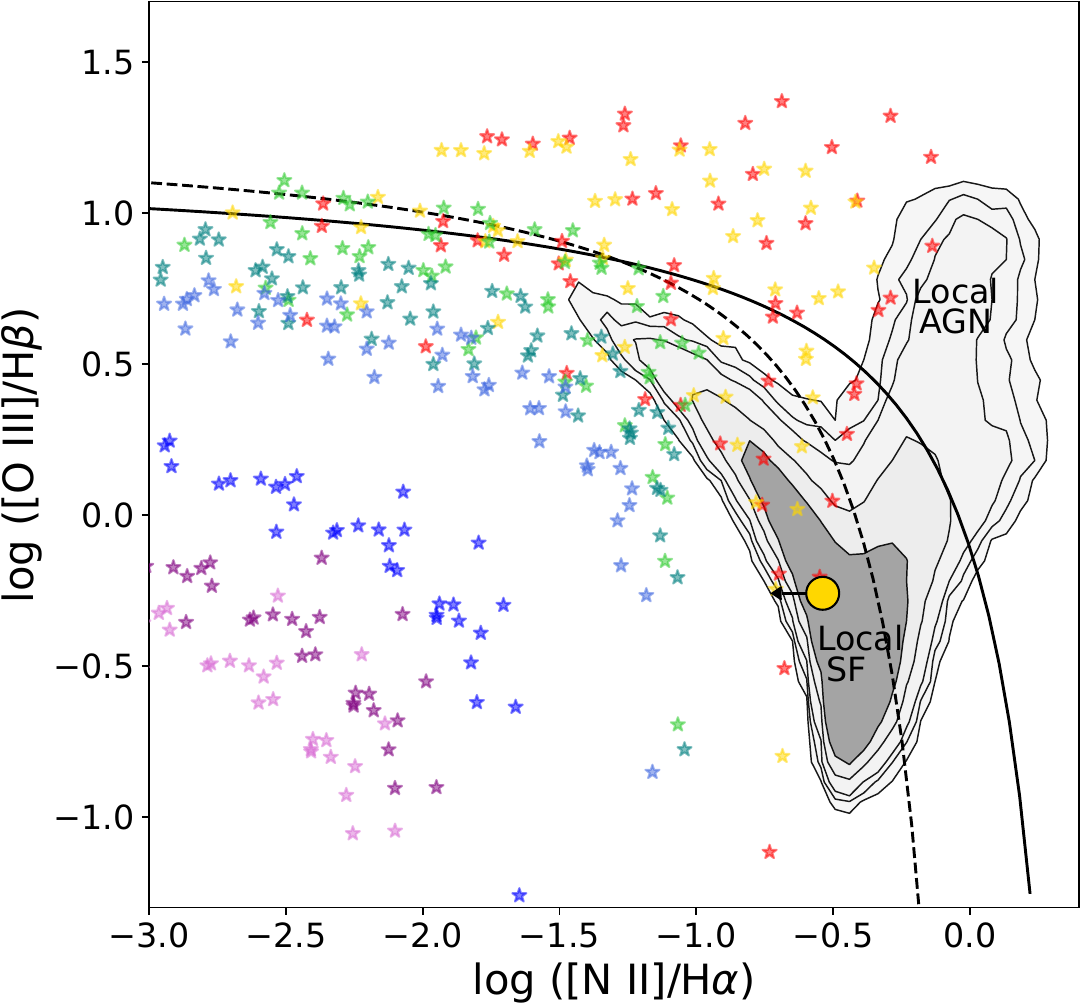"}
\includegraphics[height=0.35\linewidth]{"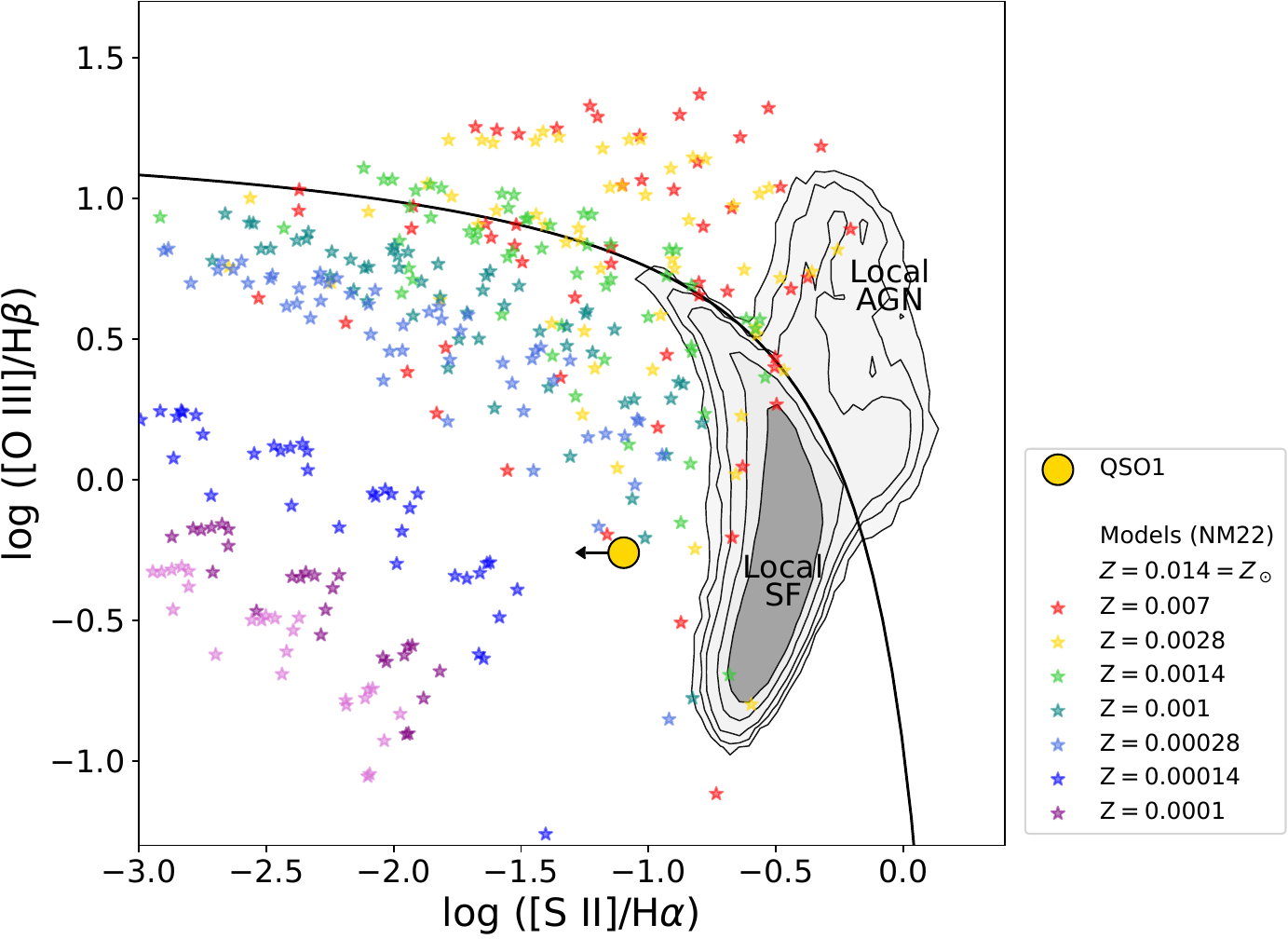"}
	\caption{
    BPT emission line ratio diagnostic 
    diagrams\citet{bpt}. Left: log([OIII]5007/H$\beta$) vs log([NII]6584/H$\alpha$). Right: log([OIII]5007/H$\beta$) vs log([SII]6716,6731/H$\alpha$). In both diagrams contours show the distribution of local galaxies. The solid black line indicates the limit for maximal starburst identified by \citet{Kewley2001}, while the black dashed line is the local demarcation between star forming galaxies and AGN identified by \citet{Kauffmann2003}.
     Small starred symbols are AGN photoionization models from \citet{Nakajima2022} spanning different densities and ionization parameters, and color-coded by metallicity.
    The solid golden circle is the observation of QSO1 (central region), for which only relatively loose upper limits are available on [NII]/H$\alpha$ and [SII]/H$\alpha$, although the latter already excludes most high metallicity models.
	}
	\label{fig:bpt}
\end{figure*}

In principle having information on the [OIII]/H$\beta$ 
we could locate QSO1 on the classical excitation diagnostic diagrams, such as the \cite{bpt}. These diagrams have turned out to be much less discriminatory of the galaxy properties at high redshift, and in particular the distinction between AGN and star forming galaxies. Indeed, at z$>$4 AGN have found to mostly overlap with star forming galaxies, an effect which has mostly been ascribed to the lower metallicity of the NLR \citep[e.g.][]{Ubler2023,Maiolino_AGN,Juodzbalis2025type1JADES}, an issue that has prompted the development of alternative diagnostics based on fainter optical lines or UV lines \citep[e.g.][]{Scholtz2025_type2,Mazzolari2024}. In the case of QSO1 the BPT diagarms are even more problematic as the low ionization lines that are required to locate the source are not detected, and the upper limit are not very informative, as discussed in the following.

The left panel of Fig.\ref{fig:bpt} shows the so-called [NII]-BPT diagram, i.e. [OIII]5007/H$\beta_N$ versus [NII]6584/H$\alpha _N$, where contours show the distribution of local galaxies and the solid black line indicates the limit for maximal starburst identified by \citet{Kewley2001}, while the black dashed line is the local demarcation between star forming galaxies and AGN identified by \citet{Kauffmann2003}.
\cite{DEugenio2025} provides an upper limit of 0.29 on the ratio [NII]/H$\alpha$ in QSO1 (Tab. \ref{tab:fluxes}). This upper limit, together with our the [OIII]/H$\beta$ ratio inferred in our work, is shown with a golden circle. The starred small symbols are from the AGN photoionization models presented in \cite{Nakajima2022}, which span a wide range of ionization parameters, ionizing spectrum shapes, and color coded by metallicity as indicated in the legend. Clearly, the current upper limit on [NII]/H$\alpha$ is not very constraining, except for indicating that QSO1 is not located in the local AGN locus, as is the case for many other AGN at high-z.

The right panel of Fig.\ref{fig:bpt} shows the so-called [SII]-BPT diagram, where [NII] is replaced by the flux of the [SII]6717,6731 doublet. In this case we report the [SII]/H$\alpha$ inferred in our work (Tab.\ref{tab:fluxes} and Appendix \ref{app:low_ion}). In this case the location of QSO1 is somewhat more constraining. It indicates tha the location of QSO1 is not only inconsistent with local AGN, but also inconsistent with the locus of local star forming galaxies. Additionally, when compared with photoionization models by \cite{Nakajima2022}, it indicates that  it is mostly inconsistent with metallicities higher than $\sim 2.8 \times 10^{-3}$, i.e. $Z<2\times 10^{-2}$, although the upper limit could be marginally consistent with some rare cases of high metallicity models.

\section{Variability}
\label{sec:variability}

Because of the different travel times, the three lensed images of QSO1 are observed at different epochs. Specifically, image C
is followed by image A 18-19 years later (i.e. 2.2-2.4 years later in the rest-frame), and this is then followed by image B another 2.2-3 years later (i.e. 3.2-4.5 months later in the rest-frame). Therefore, the three images offer, in principle, the possibility of investigating the AGN variability on different timescales. However, simply comparing the fluxes of the three images (corrected for magnification) does not provide a robust  method to explore variability, due the uncertainties in the lens magnification of each of the three images. 
Yet, variability between the three images can be explored in terms of EW of the emission lines, as this quantity is independent of the magnification factor. Indeed,
both \cite{Ji2025} and \cite{Furtak2025variab} have reported that the EW of H$\beta$ and H$\alpha$ in image C of QSO1 is significantly larger than in images A and B. \cite{Ji2025} interpret this variation in terms of a decrease of the accretion disc continuum to which the BLR has not yet reacted, because of time delay; this interpretation is also supported by the lower continuum observed in the spectroscopic observation relative to photometric data of the same image obtained one year earlier.

Having evidence that QSO1 is variable, one may wonder whether variability may play a role in the observed narrow line ratios. We have seen that the narrow lines are emitted on scales of a few 100 pc, hence even if they are ionized by the AGN (and not by star formation), they are not sensitive to AGN continuum variability on timescales shorter than a few 100 years. However, it is possible that the currently observed narrow lines were excited by the AGN at an epoch when it was experiencing different rates of accretion, hence producing higher or lower ionizing radiation. While this can certainly modulate the absolute intensity of the lines, is not obvious that this could affect the line ratios; in particular, both H$\beta _N$ and [OIII] would brighten or become fainter together, proportionally. One possibility is that the dimming of the ionizing radiation may have been so extreme that could have resulted in an extremely low ionization parameter, which may have produced a very low [OIII]/H$\beta$; however, the low inoization parameter scenario is ruled out in  appendix  \ref{app:low_ion} on various grounds.

Another possibility is that the different narrow lines may react on different timescales to the long term variations of the continuum ionizing emission, possibly because of large scale stratification of the emitting regions. This has been observed in some nearby AGN \citep{Rashed2015}. However, even in the most extreme cases the differential temporal variation of the [OIII] and H$\beta$ lines is only 10--15\%, which would certainly not account for the very low [OIII]/H$\beta$ observed in QSO1, a factor of a few/several lower than other (chemically evolved) AGN at lower redshifts.

\section{Semi-Analytical Models and Hydrodynamical Simulations}
\label{app:models}

In this section we provide additional information on the semi-analytical models and hydrodynamical simulations that have been used in Fig.~\ref{fig:models}.

\subsection{DELPHI}
\label{app:models-delphi}

{\sc Delphi} ({\bf D}ark Matter and the {\bf e}mergence of ga{\bf l}axies in the e{\bf p}oc{\bf h} of re{\bf i}onization) is a semi-analytic model that uses a binary merger tree approach to jointly track the build-up of dark matter halos, their baryonic components (gas, stellar, dust and metal masses) and black holes at $z >= 4.5$  \citep{dayal2014,dayal2019,piana2022,Dayal2024DELPHI} using the Planck 2020 cosmological model \citep{Planck2020}. We follow the assembly of dark matter halos between $\log_{10}(M_{\rm h}/\rm M_\odot)=8-14$ from $z \sim 40$ down to $z = 4.5$ with a mass resolution of $10^8 \, \rm M_\odot$. Crucially, this model has been calibrated against the latest datasets for both star forming galaxies and black holes assimilated by the JWST and ALMA, and run exploring: (i) different seeding prescriptions including {\it light} ($100 \rm M_\odot$) and {\it heavy} seeds ($10^{3-5}\rm M_\odot$); (ii) different spins exploring cases with spin values of $s=0,+1,-1$; (iii) allowing black hole growth in both Eddington-limited and super-Eddington accretion scenarios. In this model, the growth of black holes is regulated both by the host halo mass and the gas mass left after star formation and the associated Type II Supernova (SNII) feedback. Our model includes a ``critical" halo mass for efficient black hole accretion with a value that evolves with redshift as $M_{\rm bh}^{\rm crit}(z) = 10^{11.25}[\Omega_m(1+z)^3 +\Omega_\Lambda ]^{-0.125}$ on which we include a scatter of 0.5 dex, motivated by the results of cosmological simulations \citep{bower2017}. In order to explain the number density of JWST-detected AGN, at any time-step black holes are allowed to accrete a gas mass of $M_{\rm bh}^{\rm crit}(z) = min[\epsilon_r f_{\rm bh}^{\rm ac}  M_{\rm g}^{\rm sf}, f_{\rm Edd}~ m_{\rm Edd} ]$ where $M_{\rm g}^{\rm sf}$ is the gas mass left after star formation and its associated SNII feedback, $f_{\rm Edd}$ is the Eddington fraction and $m_{\rm Edd}$ is the Eddington accretion rate. Allowing very weak AGN feedback ($0.01\%$ of black holes feedback coupling to the gas) we require values of $f_{\rm bh}^{\rm ac} = 0.1 ~ (5 \times 10^{-4})$ and $f_{\rm Edd} = 1.0 ~ (10^{-4})$ for halos above (below) the critical mass (we allow 0.5 dex of scatter on all of these quantities); i.e. black holess in high-mass (low-mass) halos can accrete the minimum between 10\% ($0.05\%$) of the available gas mass and $100\% ~(0.01\%)$ of the Eddington fraction. Assuming instantaneous recycling and perfect mixing of gas, dust and metals, we model the interlinked dust and metal contents of early systems including all of the relevant processes: for dust we account for production in SNII, destruction in SNII shocks, astration into star formation, ejection in outflows and grain growth in the cold inter-stellar medium. The time-evolution of metal enrichment is calculated as
\begin{equation}
\frac{d M_{\rm Z}}{dt} = \dot M_{\rm Z}^{\rm pro}  - \dot M_{\rm Z}^{\rm ast} + \dot M_d^{\rm des} -\dot M_{\rm Z}^{\rm eje} - \dot M_d^{\rm gro}.
\label{eq_met}
\end{equation}
where the terms on the right hand side account for the rate of metal production ($M_{\rm Z}^{\rm pro}$) including the latest state-of-the-art yields from Type Ia SN (SNIa), SNII and Asymptotic Giant branch (AGB) stars from \cite{kobayashi2020}, the rate of metals astrated into star formation ($\dot M_{\rm Z}^{\rm ast}$), the rate of dust mass destruction in SNIII shocks that adds to the metal content ($\dot M_d^{\rm des}$), the rate of metals lost in outflows ($\dot M_{\rm Z}^{\rm eje}$ ) and  the rate of metals lost to dust grain growth in the cold ISM ($M_d^{\rm gro}$). This model has already been used to study the contribution of black holes to the reionization process \citep{Dayal2024DELPHI} and been pushed to its extreme limits to explain the enormous black hole-to-stellar mass ratios being observed with the JWST \citep{Furtak2023_AGN}.

\subsection{CAT}
\label{app:models-cat}

The \textsc{Cosmic Archaeology Tool} \citep{Trinca2022,Trinca2023} is a versatile semi-analytical model designed to trace and reproduce the early galaxy evolution during the first Gyr of cosmic history. Its primary goal is to explore how different formation and accretion scenarios for the first BH seeds contribute to the build-up of the massive BH population, as well as their co-evolution with host galaxies over cosmic time. CAT runs on a suite of semi-analytical dark matter merger trees, offering a large statistical sampling of the galaxy population from $z \approx 25$ down to $z = 4$. It resolves the formation of the first cosmic structures, with a minimum halo mass resolution of $\sim 10^6 \, \rm M_\odot$, following the hierarchical assembly of dark matter halos up to $\rm{log}(M_{h}/M_\odot) = 14$.
In each halo, CAT characterizes the formation of both PopIII and PopII stars, depending on the galaxy metallicity \citep{Trinca2024b}. The code then self-consistently tracks the enrichment of the ISM with dust and metals from both SNe and AGB stars, relying on mass- and metallicity-dependent stellar yields.
The model include a two-phase ISM, where dust grains can be destroyed by SN shocks in the diffuse hot medium or grow in mass by accreting gas-phase metals in warm, dense gas. Mechanical feedback from SN explosions and AGN accretion is also included, potentially driving energy-driven galaxy-scale winds \citep{Valiante2016}.
The enrichment of IGM is evolved across different merger trees,  consistently with the predicted galaxy outflows. It is worth noting though that, due to the lack of spatial information on the DM halo distribution (which is an  intrinsic limitation of our semi-analytical merger trees) the IGM metallicity is modeled as an average quantity. This value evolves with redshift and depends on the simulated overdensity. In reality, the patchy IGM enrichment might translate in lower metallicity levels than those predicted by the model for systems evolving in relative isolation.

The BH formation is implemented following two different seeding channels:
\textit{light seeds} ($M_{\rm seed} \sim 100 \, M_\odot$) form as remnants of PopIII stars, (formed in galaxies with metallicity below a critical threshold of $Z_{\rm crit} = 10^{-3.8} \, Z_\odot$). When the stellar population evolves, the most massive BH remnant formed is retained as the galaxy nuclear BH.
\textit{Heavy seeds} ($M_{\rm seed} = 10^5 \, M_\odot$) are assumed to form instead through the so-called direct collapse scenario \citep{Bromm2003,Begelman2006}, under specific conditions for the host galaxy. The host halo must be metal-poor ($Z < Z_{\rm crit}$), support atomic hydrogen cooling ($T_{\rm vir} > 10^4 \, \rm K$), and be exposed to a sufficiently strong Lyman-Werner radiation to suppress molecular hydrogen formation and prevent gas fragmentation, enabling the monolithic collapse of the gas. This last condition corresponds to a critical LW flux threshold of $J_{\rm LW} = 300 \, J_{21}$, where $J_{21} = 10^{-21} \, \rm erg \, s^{-1} \, cm^{-2} \, sr^{-1} \, Hz^{-1}$.

The catalogues presented in this work consider two distinct accretion scenarios driving the subsequent BH growth \citep{Trinca2022}. In the Eddington-limited (EL) model, nuclear BHs grow at the Bondi rate \citep{Bondi1952}, with accretion capped at the Eddington limit. Due to the strong dependence of the Bondi rate on BH mass, light seeds are unable to grow significantly in this scenario. As a result, the massive BH population at high redshift originates exclusively from heavy seed progenitors. In the super-Eddington (SE) model, on top of the Bondi accretion, we allow for short episodes of enhanced accretion during major galaxy mergers (with typical durations of $\Delta t \sim 1 \, \rm Myr$ \citep{Trinca2024}). These bursts are assumed to be driven by strong gas inflows into the nuclear region, triggered by angular momentum loss during mergers. During these phases, the BH accretion rate is modelled as $\dot{M}_{\rm BH} = \epsilon_{\rm BH} M_{\rm gas}/\tau_{\rm bulge}$, i.e. proportional to the galaxy gas content and independent of BH mass. These episodes of enhanced accretion are often characterized by super-Eddington rates, particularly for low-mass BHs, enabling light seeds to grow efficiently already at early times. Therefore, in this scenario, both light and heavy seeds contribute to the build-up of the massive BH population, with rapidly growing light seeds representing a competitive channel to the direct collapse scenario.

\subsection{AESOPICA}
\label{app:models-fable}

\textsc{Aesopica} is a new suite of large-volume cosmological simulations (Koudmani et al., in prep) built upon the \textsc{Fable} galaxy formation model  \citep{Henden2018}. The \textsc{Fable} sub-grid models are largely based on the Illustris galaxy formation model  \citep{Vogelsberger2014}. Whilst the models for star formation \citep{Springel2003}, radiative cooling  \citep{Katz1996,Wiersma2009a}, and chemical enrichment  \citep{Wiersma2009} are unchanged from Illustris, the stellar feedback \citep{Vogelsberger2013} and AGN feedback \citep{Sijacki2015} models have been updated to include thermal stellar feedback and an AGN duty cycle. \textsc{Aesopica} introduces additional targeted updates for modelling the growth of infant black holes in the early Universe, exploring three key modifications to fiducial galaxy formation models: enabling efficient accretion in the low-mass regime \citep{Koudmani2022}, incorporating super-Eddington accretion, and examining a broad range of seed masses ($10^{2}~\mathrm{M_{\odot}}$ to $10^{5}~\mathrm{M_{\odot}}$) following seed evolution from early cosmic epochs ($z \sim 20$). This is achieved by lowering the halo mass threshold for seeding black holes to the smallest resolvable halo size to $3 \times 10^{9}~\mathrm{M_{\odot}}$. We note that this represents a likely optimistic seeding scenario, so our black hole occupation for resolved haloes should be seen as an upper limit.

\subsection{Phanes: an analytic formalism for primordial black holes}
\label{ssec:models-phanes}

We use the results obtained within the context of the {\sc phanes} ({\bf P}rimordial black {\bf h}oles {\bf a}ccelerating the assembly of {\bf n}ascent {\bf e}arly {\bf s}tructures) analytic formalism. This is described in detail in \cite{Dayal_Maiolino2025}. This work follows the time-evolution of galaxies seeded by primordial black holes. It purely focuses on the ``seed effect" where the Coulomb effect of a single black hole can generate an initial density fluctuation that grows through gravitational instability \citep{carr-silk2018,Escriva2024}. Black holes start accreting dark matter linearly starting at the redshift of matter-radiation equality ($z \sim 3400$) such that by $z \sim 34$, the dark matter halo starts dominating the potential. At this point, the halo is allowed to grow non-linearly using the accretion rate from state-of-the-art N-body simulations \citep{trac2015}. A halo is allowed to accrete gas - at a rate driven by the cosmological baryon-to-dark matter ratio - once its mass is sufficient to host a baryonic over-density of about 200. This gas can be accreted onto the black hole to allow its growth, and form stars. The {\sc phanes} formalism also accounts for the feedback from both Type II Supernovae (SNII) and BH accretion in determining the final gas mass at the end of any time-step; this acts as the initial gas mass for the next time-step in order to track the baryonic assembly of such early systems. Assuming perfect mixing of metals and gas, the formalism accounts for the key processes of metal production (where  the latest mass-dependent stellar yields from \citealt{kobayashi2020} are used), astration into black hole accretion and star formation, and loss in BH- and SNII-powered outflows in order to calculate the metal enrichment of these early systems.

\subsection{PBH hydrodynamical simulations}
\label{app:models-zhang}

Using cosmological N-body and hydrodynamic simulations with the \textsc{Gizmo} code~\citep{Hopkins2015MNRAS.450...53H}, we model the formation of the first galaxies in PBH-seeded halos, taking into account the accretion and feedback of an isolated PBH with $5\times 10^7\,M_{\odot}$, placed in a simulation box of comoving size $L \sim 1\ \mathrm{Mpc}/h$. This PBH mass scale could arise from an initial $\sim 10^6\,M_{\odot}$ seed, related to the $e^+e^-$ phase transition in the ultra-early Universe~\citep[e.g.][]{Carr2021PDU....3100755C}, and assuming subsequent growth through the merger of clustered PBHs in a highly biased region. To simulate the formation of structure around the PBH, we adopt the numerical recipes summarized in~\citet{Zhang2025arXiv250317585Z}, to which we refer for further details.  The simulations are run to $z \sim 7$, when QSO1 is observed. We specifically implement a star formation prescription for Jeans unstable gas in the vicinity of an accreting PBH, taking into account the intricate interaction between the different gaseous components. 

The accretion-driven growth of the PBH is sensitive to the coupling strength of the accretion-generated luminosity, $L_{BH}$, to the surrounding gas, implemented via thermal energy injection: $\delta E=\epsilon_r L_{BH} \delta t$, given a simulation timestep $\delta t$. In this case, we take a fiducial value of $\epsilon_r \sim 0.5\%$, following ~\citet{Zhang2025arXiv250317585Z}. By $z\sim 7$, feedback heating of the ambient gas suppresses efficient accretion and further growth, resulting in a final BH mass of $M_{BH}\simeq 6\times 10^7 \, M_{\odot}$ and delayed build-up of stellar mass, yielding $M_{star} \sim 2 \times 10^7 \, M_{\odot}$—both comparable to current observational estimates for QSO1.
The (radiation-hydrodynamical) gas flows in the vicinity of the accreting PBH, however, are complex, and follow-up simulations are required to determine the branching ratio of such large-scale inflows into feeding of the black hole and of the concurrent starburst, covering a statistically representative sample of environments.  

We track the metallicity evolution in galaxies by post-processing, incorporating both star formation and gas outflows driven by black hole feedback. The metallicity $Z$ of a galaxy at a given time $t$ is estimated as the ratio of metal mass $M_Z(t)$ to gas mass $M_{gas}(t)$, contained within the outflow bubble: 
\begin{equation} 
Z(t) = \frac{M_Z(t)}{M_{gas}(t)}\mbox{\ ,} 
\end{equation} 
assuming a uniform distribution within this bubble and ignoring metal diffusion at the bubble surface.

The metal mass is given by
\begin{equation}
    M_Z(t) = \mathcal{M}_{Z}M_{star}(t)\mbox{\ ,}
\end{equation}
where $M_{star}(t)$ is the stellar mass as a function of cosmic time, and $ \mathcal{M}_{Z} $ is the metal yield per unit stellar mass, which depends on the stellar initial mass function (IMF) and stellar evolution models for winds and supernovae. Based on the stellar evolution models in \cite{Costa:2025esr}, we estimate $\mathcal{M}_{Z}$ to be in the range $\sim 0.02-0.19$, bracketing the plausible range for different IMFs (from present-day to Pop~III stars).

 We estimate the size of the outflow bubble as the radius where the average outflow velocity drops below the local sound speed, and further simplify our model by neglecting the delay between star formation and subsequent metal enrichment due to stellar evolution. Fig.~\ref{fig:models}d shows a representative trajectory for the resulting metallicity evolution with the initial conditions described above. By $z \sim 7$, the metallicity is estimated to lie within $Z/Z_{\odot} \sim 0.002 - 0.02$, consistent with what is observed for QSO1. We again refer to \cite{Zhang2025arXiv250317585Z} for a detailed description of the numerical implementation and simulation parameters. We note, however, that this post-processed enrichment model remains an approximation; a more complete treatment incorporating explicit stellar evolution and metal enrichment recipes will be explored in an upcoming work.
 

\section*{Authors' affiliations}
\noindent
$^{1}$Kavli Institute for Cosmology, University of Cambridge, Madingley Road, Cambridge CB3 0HA, UK
\\
$^{2}$Cavendish Laboratory, University of Cambridge, 19 JJ Thomson Avenue, Cambridge CB3 0HE, UK
\\
$^{3}$Department of Physics and Astronomy, University College London, Gower Street, London WC1E 6BT, UK
\\
$^{4}$Max-Planck-Institut für extraterrestrische Physik, Gießenbachstraße 1, 85748 Garching, Germany
\\
$^{5}$Centro de Astrobiolog\'ia (CAB), CSIC–INTA, Cra. de Ajalvir Km.~4, 28850- Torrej\'on de Ardoz, Madrid, Spain
\\
$^{6}$Department of Astronomy, University of Texas at Austin, Austin, TX 78712, USA
\\
$^{7}$Weinberg Institute for Theoretical Physics, Texas Center for Cosmology and Astroparticle Physics,
University of Texas at Austin, Austin, TX 78712, USA
\\
$^{8}$Canadian Institute for Theoretical Astrophysics, 60 St George St, University of Toronto, Toronto, ON M5S 3H8, Canada
\\
$^{9}$St Catharine’s College, University of Cambridge, Trumpington Street, Cambridge CB2 1RL, UK
\\
$^{10}$Institute of Astronomy, University of Cambridge, Madingley Road, Cambridge, CB3 0HA, UK
\\
$^{11}$Center for Computational Astrophysics, Flatiron Institute, 162 Fifth Avenue, New York, NY 10010, USA
\\
$^{12}$Centre for Astrophysics Research, Department of Physics, Astronomy and Mathematics, University of Hertfordshire, College Lane, Hatfield, AL10 9AB, UK
\\
$^{13}$
Universit\"{a}t Heidelberg, Zentrum f\"{u}r Astronomie, Institut f\"{u}r Theoretische Astrophysik, D-69120 Heidelberg, Germany
\\
$^{14}$Dipartimento di Fisica, ‘Sapienza’ Universit\`
a di Roma, Piazzale Aldo Moro 2, I-00185 Roma, Italy
\\
$^{15}$INAF/Osservatorio Astronomico di Roma, Via di Frascati 33, I-00040 Monte Porzio Catone, Italy
\\
$^{16}$INFN, Sezione Roma1, Dipartimento di Fisica, ‘Sapienza’ Universit\`
a di Roma, Piazzale Aldo Moro 2, I-00185 Roma, Italy
\\
$^{17}$Sapienza School for Advanced Studies, Viale Regina Elena 291, I-00161 Roma, Italy
\\
$^{18}$
Como Lake Center for Astrophysics, DiSAT, Universit\`
a degli Studi dell’Insubria, via Valleggio 11, 22100, Como, Italy
\\
$^{19}$Sorbonne Universit\'e, CNRS, UMR 7095, Institut d'Astrophysique de Paris, 98 bis bd Arago, 75014 Paris, France
\\
$^{20}$Kavli Institute for Astronomy and Astrophysics, Peking University, Beijing 100871, China
\\
$^{21}$Scuola Normale Superiore, Piazza dei Cavalieri 7, I-56126 Pisa, Italy
\\
$^{22}$Institute of Liberal Arts and Science
Kanazawa University
Kakuma
Kanazawa, Ishikawa 
920-1192, JAPAN
\\
$^{23}$Waseda Research Institute for Science and Engineering, Faculty of Science and Engineering, Waseda University, 3-4-1, Okubo, Shinjuku, Tokyo 169-8555, Japan
\\
$^{24}$Cosmic Dawn Center (DAWN), Copenhagen, Denmark
\\
$^{25}$Niels Bohr Institute, University of Copenhagen, Jagtvej 128, DK-2200, Copenhagen, Denmark
\\
$^{26}$Department of Physics, University of Oxford, Denys Wilkinson Building, Keble Road, Oxford OX1 3RH, UK
\\
$^{27}$European Southern Observatory, Karl-Schwarzschild-Strasse 2, 85748 Garching, Germany
\\
$^{28}$
Universit\`a di Firenze, Dipartimento di Fisica e Astronomia, via G. Sansone 1, 50019 Sesto Fiorentino, Florence, Italy 2
\\
$^{29}$INAF – Arcetri Astrophysical Observatory, Largo E. Fermi 5, I50125, Florence, Italy
\\
$^{30}$Department of Astronomy \& Astrophysics, University of Chicago, 5640 S Ellis Avenue, Chicago, IL 60637, USA
\\
$^{31}$Kavli Institute for Cosmological Physics, University of Chicago, Chicago IL 60637, USA
\\
$^{32}$AURA for European Space Agency, Space Telescope Science Institute, 3700 San Martin Drive. Baltimore, MD 21210, USA
\\
$^{33}$Aix Marseille Universit\'e, CNRS, CNES, LAM (Laboratoire d’Astrophysique de Marseille), UMR 7326, F-13388 Marseille, France
\\
$^{34}$Department of Astronomy and Astrophysics, University of California, Santa Cruz, 1156 High Street, Santa Cruz, CA 96054, USA 
\\
$^{35}$Center for Astrophysics $|$ Harvard \& Smithsonian, 60 Garden St., Cambridge MA 02138, USA

\bsp	
\label{lastpage}
\end{document}